\documentclass{aa}
\usepackage{graphicx}
\usepackage{multirow}
\usepackage{txfonts}
\usepackage{xspace}
\usepackage{xcolor}
\usepackage{url}
\usepackage[colorlinks=true, allcolors=blue]{hyperref}
\usepackage[flushleft]{threeparttable}
\usepackage{orcidlink}
\newcommand{\ixpe}{\text{IXPE}\xspace}

\begin{document}

\title{Probing the polarized emission from the accretion-powered pulsar 4U 1907+09 with \ixpe}

\author{Menglei Zhou \inst{\ref{in:Tub}}\orcidlink{0000-0001-8250-3338}
\and Lorenzo Ducci \inst{\ref{in:Tub},\ref{in:ISDC}}\orcidlink{0000-0002-9989-538X}
\and Honghui Liu \inst{\ref{in:Tub}\orcidlink{0000-0003-2845-1009}}
\and Sergey~S.~Tsygankov \inst{\ref{in:UTU}}\orcidlink{0000-0002-9679-0793}
\and Sofia~V.~Forsblom \inst{\ref{in:UTU}}\orcidlink{0000-0001-9167-2790}
\and Alexander~A.~Mushtukov \inst{\ref{in:Oxford}}\orcidlink{0000-0003-2306-419X}
\and Valery~F.~Suleimanov \inst{\ref{in:Tub}}\orcidlink{0000-0003-3733-7267}
\and Juri Poutanen \inst{\ref{in:UTU}}\orcidlink{0000-0002-0983-0049}
\and Pengju Wang \inst{\ref{in:Tub}}\orcidlink{0000-0002-6454-9540}
\and Alessandro Di Marco \inst{\ref{in:INAF-IAPS}}\orcidlink{0000-0003-0331-3259}
\and Victor Doroshenko \inst{\ref{in:Tub}}\orcidlink{0000-0001-8162-1105}
\and Fabio La Monaca \inst{\ref{in:INAF-IAPS}, \ref{in:UniRoma2}}\orcidlink{0000-0001-8916-4156}
\and Vladislav Loktev \inst{\ref{in:UTU}, \ref{in:Hel}}\orcidlink{0000-0001-6894-871X}
\and Alexander Salganik \inst{\ref{in:UTU}}\orcidlink{0000-0003-2609-8838}
\and Andrea Santangelo \inst{\ref{in:Tub}}\orcidlink{0000-0003-4187-9560}
}

\institute{
Institut f\"ur Astronomie und Astrophysik, Universit\"at T\"ubingen, Sand 1, D-72076 T\"ubingen, Germany \label{in:Tub} \\ \email{menglei.zhou@astro.uni-tuebingen.de}
\and
ISDC Data Center for Astrophysics, Universit\'e de Gen\`eve, 16 chemin d'\'Ecogia, 1290 Versoix, Switzerland \label{in:ISDC}
\and Department of Physics and Astronomy, 20014 University of Turku, Finland \label{in:UTU}
\and Astrophysics, Department of Physics, University of Oxford, Denys Wilkinson Building, Keble Road, Oxford OX1 3RH, UK \label{in:Oxford}
\and INAF Istituto di Astrofisica e Planetologia Spaziali, Via del Fosso del Cavaliere 100, 00133 Roma, Italy \label{in:INAF-IAPS}
\and 
Dipartimento di Fisica, Universit\`{a} degli Studi di Roma ``Tor Vergata'', Via della Ricerca Scientifica 1, 00133 Roma, Italy \label{in:UniRoma2} 
\and Department of Physics, P.O. Box 64, 00014 University of Helsinki, Finland \label{in:Hel}
}

\titlerunning{Probing the polarized emission from the accretion-powered pulsar 4U~1907+09 with \ixpe}
\authorrunning{M.~Zhou et al.}

\date{-- / --}

\abstract{We present observations of the accretion-powered X-ray pulsar 4U~1907+09 conducted with the Imaging X-ray Polarimetry Explorer, which has delivered the first high-quality polarization measurements of this source. 4U~1907+09 was observed twice during its brightest periods, close to the periastron. 
We observe a stronger polarization in the first observation, with a phase-averaged polarization degree (PD) of $6.0 \pm 1.6\%$ and a polarization angle (PA) of $69\degr \pm 8\degr$. The second observation provides weaker constraints on the polarimetric properties, PD=$2.2 \pm 1.6\%$ and PA=$46\degr \pm 23\degr$, as determined from the spectro-polarimetric analysis. Combining the data from the two observations results in PD=$3.7 \pm 1.1\%$ and PA=$63\degr \pm 9\degr$. We detect an energy-dependent PA in the phase-averaged analyses with a significance of 1.7\,$\sigma$. In the phase-resolved analyses, we observe a potential PA rotation of approximately 90\degr\ between adjacent energy bands (4--5 and 5--6\,keV) within the single phase bin of 0.25--0.375. We also investigate the influence of short flares on the polarization properties of this source. The results suggest that flares do not significantly affect the energy-phase-dependent PA, implying that the pulsar's geometry remains stable during flare events. 
}


\maketitle

\section{Introduction}\label{Sect:intro}

High-mass X-ray binaries (HMXBs) comprise a compact object that accretes matter from a massive companion star. In HMXB systems with strongly magnetized neutron stars (NSs), the accretion flow is interrupted by the NS's magnetosphere at a distance of approximately $10^{8}$--$10^{9}$\,cm \citep{Lamb_1973, Elsner_1977}. The accreted matter is then channeled along magnetic field lines onto the NS's surface, where its gravitational potential energy is released. Depending on the accretion rate, the accretion process can generate hot spots or extended accretion columns at the magnetic poles. Due to the misalignment between the magnetic poles and the spin axis, the system emits pulsed X-rays, manifesting as an X-ray pulsar (XRP; see, e.g., \citealt{Mushtukov_Tsygankov_2024_Review} for a recent review). XRPs serve as natural laboratories for studying the extreme conditions where matter, light, and magnetic fields interact in ways unattainable in terrestrial labs. 

The phase-dependent characteristics of pulsed X-ray emission in XRPs are determined by the geometry of the emission region, the orientation of the pulsar relative to the observer, and the intricacies of radiative transfer processes in the presence of an intense magnetic field. Despite this understanding, a comprehensive theoretical framework capable of fully describing these properties is lacking. One of the primary challenges is the uncertainty surrounding the fundamental geometric configuration of XRPs. 

Polarimetric observations provide a means to address this uncertainty, by enabling the determination of key geometric parameters, such as the inclination of the pulsar's spin axis with respect to the line of sight and the magnetic obliquity. They also offer the potential to distinguish between different emission region geometries. 

The NS's extreme magnetic field significantly influences X-ray emission polarization from XRPs by modifying the Compton scattering cross sections. Under favorable orientations, photon scattering in such a highly magnetized plasma is predicted to result in a substantial degree of polarization, reaching up to 80\% \citep{Meszaros_1988_RVM, Caiazzo_2021}. As shown by \citet{Meszaros_1988_RVM}, the linear polarization of X-rays is highly sensitive to the geometry of the emission region and exhibits variability with energy and pulse phase. Consequently, phase-resolved polarimetry serves as a critical tool for constraining the system's viewing geometry and for differentiating between competing radiation models. 

\begin{figure*}
\centering
\includegraphics[width=0.95\textwidth]{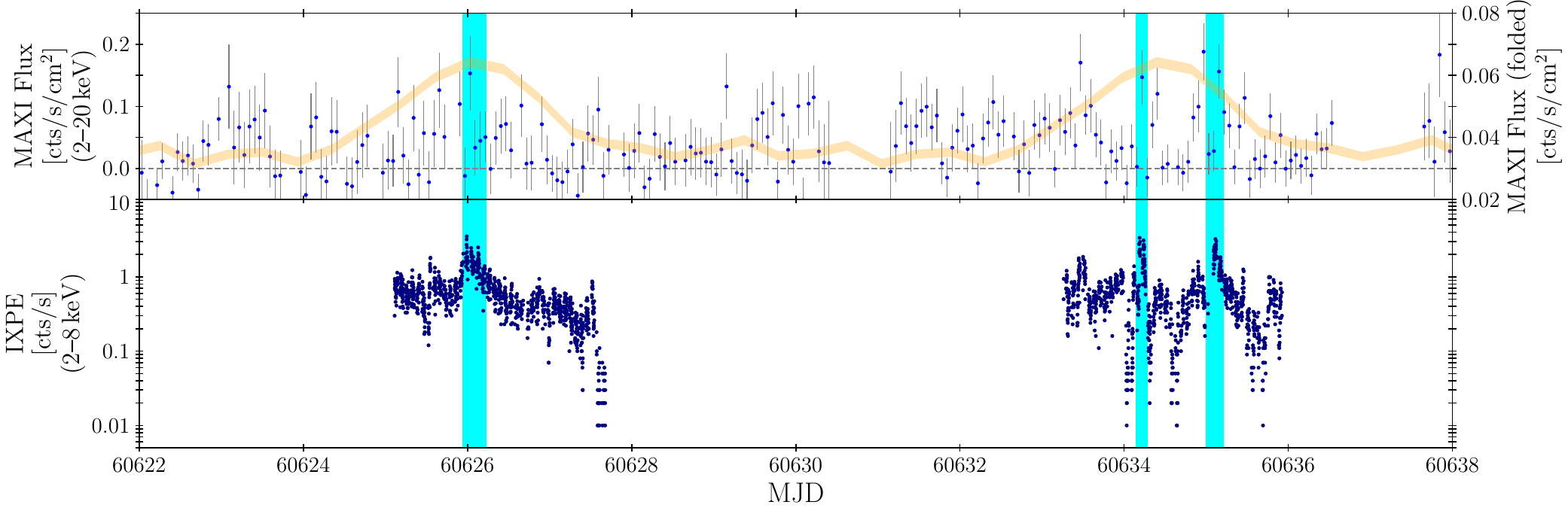}
\caption{X-ray light curve of 4U~1907+09. Upper panel: Light curve in the 2--20\,keV band as observed by MAXI near the IXPE observations. We also present the folded orbital light curve for MAXI data after MJD\,59000 based on the ephemerides from \citet{intZand_1998}, displayed in light orange, with its width representing flux uncertainties. Note that the folded orbital light curve uses the right-hand axis to emphasize its shape. Lower panel: Combined light curve of 4U~1907+09 from all three DUs of IXPE, with a time resolution of 100\,s. The duration of the three flares observed by \ixpe is indicated by cyan stripes. }
\label{fig:longterm-lc}
\end{figure*}

The Imaging X-ray Polarimetry Explorer (\ixpe) is the first mission dedicated to imaging X-ray polarimetry, enabling the detection of linear X-ray polarization in numerous astrophysical X-ray sources, including XRPs (see \citealt{Poutanen_2024_Galaxies} for a review). Most studied pulsars show no significant dependence of the polarization angle (PA) on energy, with Vela X-1 and 4U~1538$-$52 being the only known exceptions~\citep{Forsblom_2023_VelaX1, Loktev_2025_4U1538}. However, they exhibit a polarization degree (PD) that is significantly lower than any theoretical predictions. 

The source reported in this work, 4U~1907+09, was first detected in the third Uhuru survey \citep{Giacconi_1971, Forman_1978} and subsequently confirmed as an HMXB with an O-type supergiant donor star \citep{vanKerkwijk_1989, Cox_2005, Nespoli_2008}. It hosts a pulsar with a spin period of $\sim 437.5$\,s~\citep{Makishima_1984} orbiting a highly reddened companion in an eccentric orbit ($e\simeq$ 0.28) with a
period of $\simeq$ 8.3753\,d~\citep{Marshall_1980, intZand_1998}. \citet{Cox_2005} estimated a lower limit on the distance ($d$) of 5\,kpc, while \citet{Nespoli_2008} reported $d \simeq 2.8$\,kpc. More recently, \citet{Bailer-Jones_2021} derived $d \simeq 1.9$\,kpc using \textit{Gaia} Early Data Release 3. 

The broadband X-ray spectrum of 4U~1907+09 can be modeled by an absorbed power law with a high energy cutoff \citep{Ferrigno_2022}. The cyclotron resonance scattering feature (CRSF) at 17--19\,keV, along with its harmonic at approximately 38\,keV, has been reported by \citet{Mihara_1995}, \citet{Cusumano_2000}, and \citet{Tobrej_2023}. Moreover, \citet{Tobrej_2023} report an absorption line feature at $\sim$8\,keV, which might be produced by the K$\beta$ transition of \ion{Fe}{xxv}. Based on the presence of the CRSF, the magnetic field strength of 4U~1907+09 is estimated to be $2 \times 10^{12}$~G~ \citep[see, e.g.,][and references therein]{Hemphill_2013, Varun_2019}. However, the CRSFs were not detected in \textit{Rossi} X-ray Timing Explorer or \textit{Suzaku} data \citep{intZand_1998, Fuerst_2012}. As a highly magnetized NS, the compact object in 4U~1907+09 exhibits a complex pulse profile that depends on both energy and luminosity \citep{Mukerjee_2001, Rivers_2010, Sahiner_2012}. 

\ixpe observations, supported by high-quality data, have provided the first robust detection of the polarization properties of 4U~1907+09. The remainder of this paper is structured as follows: In Sect.~\ref{Sect:data} we present the long-term behaviors of the source and an overview of the data we analyzed. The results are presented in Sect.~\ref{Sect:results}. We discuss the results of our analysis in Sect.~\ref{Sect:discussion}. A summary and outlook are presented in Sect.~\ref{Sect:summary}.

\section{Observations and data reduction}\label{Sect:data}
\ixpe is a joint mission of NASA and the Italian Space Agency launched by a Falcon 9 rocket on 2021 December 9, intending to provide imaging polarimetry over the 2--8\,keV energy range~\citep{Weisskopf_2022}. \ixpe consists of three identical grazing incidence telescopes. Each telescope comprises an X-ray mirror assembly and a polarization-sensitive detector unit (DU) equipped with a gas-pixel detector~\citep{Soffitta_2021, Baldini_2021}. 
In addition to recording the sky coordinates, arrival time, and energy of each photon, the system also measures the scattering direction of the photo-electron, thereby enabling polarimetric analysis. 

\begin{table}
\centering
\caption{Orbital parameters for 4U~1907+09 (adopted from \citealt{intZand_1998}).} 
\begin{tabular}{lcc}
\hline
\hline
Parameter & Value & Unit\\
\hline
        Orbital period  & $8.3753^{+0.0003}_{-0.0002}$ & d  \\
        $T_{\pi/2}$  & $50134.76^{+0.16}_{-0.20}$ & MJD  \\
        $a_{\rm X}\sin i$  & $83 \pm 4$ & light-sec  \\
        Longitude of periastron  & $330 \pm 20$ & deg   \\
        Eccentricity  & $0.28^{+0.10}_{-0.14}$ &    \\
\hline
\end{tabular}
\label{table:orb-pars}
\end{table}

\begin{table} 
\centering
\caption{Pulsar ephemerides for Obs.~1 and 2 of 4U~1907+09 with IXPE. }
\begin{tabular}{lcc}
\hline
\hline
 Parameter & Obs. 1 & Obs. 2\\
\hline
        Start [MJD] & 60625.1 & 60633.3  \\
        End [MJD] & 60627.7 & 60635.9 \\
        Exposure [ks] & 117.9 & 118.0 \\
        $t_0$ [MJD] & \multicolumn{2}{c}{60625.1139919\tablefootmark{$\text{a}$}} \\
        Pulse period  [$\mathrm{s}$] & \multicolumn{2}{c}{$443.73_{-0.01}^{+0.03}$} \\
        Pulse period derivative [$10^{-8}\mathrm{s \, s^{-1}}$] &  \multicolumn{2}{c}{$5_{-6}^{+3}$} \\
\hline  
\end{tabular}
\tablefoot{\tablefoottext{a}{The reported reference time ($t_0$) defines the zero phases of the pulse profiles (see Fig.~\ref{fig:phaseogram}). }}
\label{table:time-pars}
\end{table}

\begin{figure}
\centering
\includegraphics[width=0.95\linewidth]{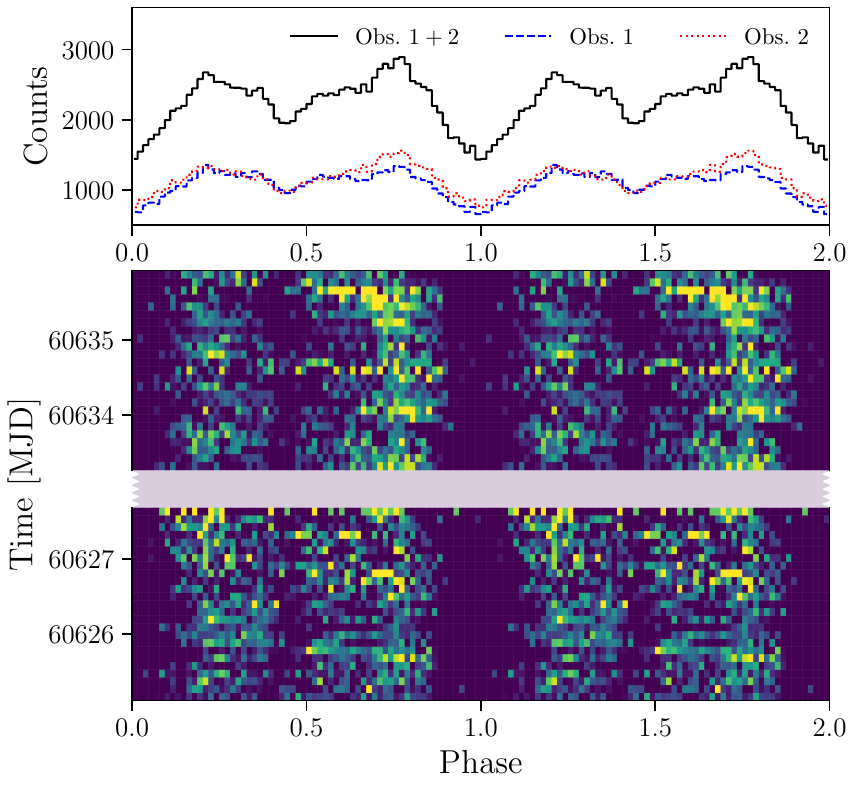}
\caption{Pulse profiles of 4U~1907+09 during two IXPE observations and their phaseograms versus time.}
\label{fig:phaseogram}
\end{figure}

\ixpe observed 4U~1907+09 twice, 2024 November 11--13 and 19--21 (hereafter Obs.~1 and 2) with total effective exposures of 117.9\,ks and 118.0\,ks, respectively, during the brightest period near the periastron (see the upper panel of Fig.~\ref{fig:longterm-lc}). 
Its light curve in the 2--8\,keV energy range from the \ixpe observatory is shown in the lower panel of Fig.~\ref{fig:longterm-lc}. 
The data have been processed using the {\sc ixpeobssim} package (version 31.0.3, \citealt{Baldini_2022_ixpeobssim}) with calibration files from the calibration database released on 2024 July 1. Before the analysis, position offset corrections and energy calibration were applied. Source photons are extracted from a circular region with a radius of $R_{\rm src}=70\arcsec$, centered on the source position. The background region is defined as an annular area centered on the source, with an inner radius of $R_{\rm in}=120\arcsec$ and an outer radius of $R_{\rm out}=240\arcsec$. The source photon flux in the 2--8\,keV band is $\sim 0.6\,\mathrm{cps}$ when combining all the three DUs, while the average background count rate in the same band is estimated at $0.001\,\mathrm{cps}\ \mathrm{arcmin}^{-2}$ per DU. Thus, the background is not subtracted when using the \texttt{pcube} algorithm provided by the {\sc ixpeobssim} team~\citep{DiMarco_2023}. However, the background effect is accounted for in the spectro-polarimetric analysis conducted with {\sc xspec}. We performed the unweighted analysis approach, following the methodology described in \citet{DiMarco_2022}. 

Event arrival times were corrected to the Solar System barycenter using the \texttt{barycorr} tool from the {\sc ftools} package. To account for the effects of binary orbital motion, additional corrections were applied using the orbital ephemerides provided by \citet{intZand_1998} extrapolated to the \ixpe observation epoch (see Table~\ref{table:orb-pars}). 

The pulse period ($P$) and its first derivative ($\dot{P}$) are independently determined by collecting photon events from both observations using $Z^2$ statistics. The obtained spin parameters and pulse epochs are listed in Table~\ref{table:time-pars}. This method naturally ensures alignment of the pulse profiles over long periods, as shown in Fig.~\ref{fig:phaseogram}. The zero point of the pulse profile is defined at the phase corresponding to the minimum count value. 

The Stokes $I$, $Q$, and $U$ spectra from all three DUs are produced via the unweighted \texttt{PHA1}, \texttt{PHA1Q}, and \texttt{PHA1U} algorithms provided in the \texttt{xpbin} tool in the {\sc ixpeobssim} package. Later, the $I$, $Q$, and $U$ spectra are re-binned to ensure at least 30 counts per energy channel, with consistent energy binning applied across all spectra. The spectra are simultaneously fitted using the {\sc xspec} package (version 12.14.0, \citealt{Arnaud_1996}) and \ixpe response matrices (ixpe:obssim20240701:v13). Fits are performed using $\chi^2$ statistics, and uncertainties are reported at the 68.3\% confidence level (1$\sigma$) unless otherwise specified. 

We also employed the \texttt{pcube} method provided by the {\sc ixpeobssim} package to perform a model-independent polarimetric analysis. This method disregards the full energy response matrix but accounts for energy dependence by weighting individual photon contributions. All photons detected within the selected time interval and energy band are used to compute the Stokes parameters through weighted summation~\citep[see Eq.~11 in][]{Baldini_2022_ixpeobssim}. 

\section{Results} \label{Sect:results}
\subsection{Light curves and pulse profiles}

4U~1907+09 was observed two separate times in the period 2024 November 11--21, corresponding to the brightest periods close to the periastron as seen in the light curve (see the upper panel of Fig.~\ref{fig:longterm-lc}) measured by MAXI\footnote{\url{http://maxi.riken.jp/}}~\citep{Matsuoka_2009_MAXI}. Three short flares, each lasting several hours, were captured during the \ixpe observations, as indicated by the cyan stripes in Fig.~\ref{fig:longterm-lc}. The fraction of photons obtained during the flares accounts for approximately 30\% of the total number of photons. 

The pulse period and pulse period derivative are measured simultaneously by maximizing the $Z^2$ statistics using all the photon events recorded within the 2--8\,keV band in both observations~\citep{Buccheri_1983}. The determined values can be found in Table~\ref{table:time-pars}. 
The resulting pulse profiles for the observations of 4U~1907+09 in the 2--8\,keV energy band are shown in Fig.~\ref{fig:phaseogram} together with the phaseograms, which display the evolution of the pulse profiles over 11 days. 

\begin{figure}
\centering
\includegraphics[width=0.9\linewidth]{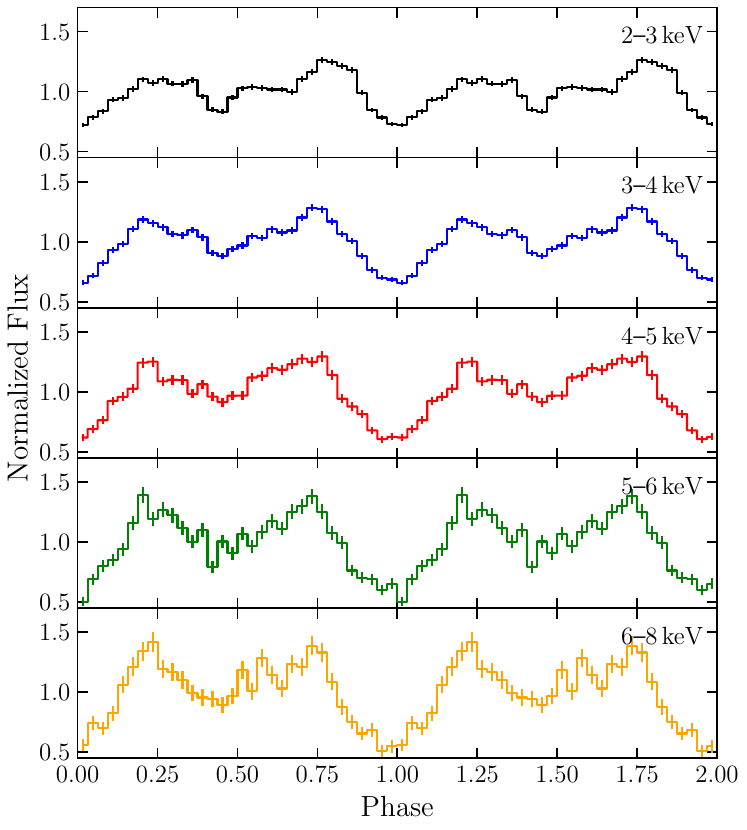}
\caption{Pulse profiles of 4U~1907+09 in different energy bands for the combined dataset (Obs.~1 and 2). The pulse profile fluxes are normalized such that their mean values are unity. }
\label{fig:pp-energy}
\end{figure}

\begin{table*}[]
\centering
\caption{Spectral parameters for the best-fit model obtained from the phase-averaged spectro-polarimetric analysis with \textsc{xspec} for Obs.~1 and 2.}
\begin{tabular}{llcccc}
\hline \hline
Component & Parameter & Unit & Obs. 1 & Obs. 2 & Obs.~1 \& 2 \\
\hline
\texttt{tbabs} & $N_{\mathrm{H}}$ & $10^{22}\mathrm{\;cm^{-2}}$ & $4.86 \pm 0.17$ & $5.12 \pm 0.17$ & $4.99 \pm 0.12$ \\
\texttt{powerlaw} & $\Gamma$ &  & $1.18 \pm 0.04$ & $1.27 \pm 0.04$ & $1.22 \pm 0.03$  \\
\texttt{polconst} & PD & \% & $6.0 \pm 1.6$ & $2.2 \pm 1.6$ & $3.7 \pm 1.1$ \\
 & PA & deg & $69 \pm 8$ & $46 \pm 23$ & $63 \pm 9$ \\
\texttt{constant} & $\mathrm{const_{DU2}}$ &  & $1.025 \pm 0.010$ & $1.027 \pm 0.009$ & $1.027 \pm 0.007$ \\
 & $\mathrm{const_{DU3}}$ &  & $1.015 \pm 0.010$ & $0.995 \pm 0.009$ & $1.005 \pm 0.007$ \\
\hline
 & $\mathrm{Flux_{2-8\;keV}}$\tablefootmark{\textbf{a}} & $10^{-10}$\,erg\,s$^{-1}$\,cm$^{-2}$  & $1.39 \pm 0.01$ & $1.47 \pm 0.01$ & $1.436 \pm 0.009$ \\
 & $\chi^2$/d.o.f. &  & 1133.1/1112 & 1201.0/1118 & 1295.9/1241 \\
\hline
\end{tabular}
\tablefoot{
The uncertainties are given at the 68.3\% (1$\sigma$) confidence level and were obtained using the \texttt{error} command in \textsc{xspec} with $\Delta\chi^2=1$ for one parameter of interest. 
\tablefoottext{\textbf{a}}{Observed (absorbed) flux in the 2--8\,keV range. } }
\label{table:phase_avg}
\end{table*}

\begin{table*} 
\centering
\caption{Polarimetric parameters in different energy bins for the two IXPE observations derived using the \textsc{xspec} method. }
\begin{tabular}{cccccccc}
    \hline\hline
      & \multicolumn{2}{c}{Obs.~1} & \multicolumn{2}{c}{Obs.~2} & \multicolumn{2}{c}{Obs.~1 \& 2} \\
    \hline
    Energy & $\text{PD}_\textsc{xspec}$ & $\text{PA}_\textsc{xspec}$ & $\text{PD}_\textsc{xspec}$ & $\text{PA}_\textsc{xspec}$ & $\text{PD}_\textsc{xspec}$ & $\text{PA}_\textsc{xspec}$ \\
    (keV) & (\%) & (deg) & (\%) & (deg) & (\%) & (deg) \\
    \hline
    2--3 & $7.1 \pm 3.4$ & $61 \pm 14$ & $3.8 \pm 3.3$ & $54 \pm 30$ & $5.4 \pm 2.4$ & $59 \pm 13$ \\
    3--4 & ${1.2}_{-1.2}^{+2.9}$ & $-55$\tablefootmark{$\dagger$} & $4.1 \pm 2.8$ & $-47 \pm 22$ & $2.6 \pm 2.0$ & $-49 \pm 26$ \\
    4--5 & $12.9 \pm 3.6$ & $74 \pm 8$ & $6.6 \pm 3.5$ & $48 \pm 16$ & $8.5 \pm 2.5$ & $65 \pm 8$ \\
    5--6 & $6.8 \pm 4.7$ & $80 \pm 22$ & ${4.1}_{-4.1}^{+4.6}$ & $-39$\tablefootmark{$\dagger$} & ${2.5}_{-2.5}^{+3.3}$ & $-78$\tablefootmark{$\dagger$} \\
    6--8 & $13.1 \pm 5.7$ & $49 \pm 13$ & $17.0 \pm 5.6$ & $37 \pm 10$ & $13.4 \pm 4.0$ & $43 \pm 9$ \\
    \hline
    2--8 & $6.0 \pm 1.6$ & $69 \pm 8$ & $2.2 \pm 1.6$ & $46 \pm 23$ & $3.7 \pm 1.1$ & $63 \pm 9$\\      
    \hline
    \end{tabular}
\tablefoot{The uncertainties computed using the \texttt{error} command 
are given at the 68.3\% (1$\sigma$) confidence level ($\Delta\chi^2 = 1$ for one parameter of interest). \\
\tablefootmark{$(\dagger)$}{The PA cannot be confined within the 1$\sigma$ confidence level; therefore, we present the most probable value instead. } } 
\label{table:pol_ebins_xspec}
\end{table*}

\begin{table*} 
\centering
\caption{Polarimetric parameters in different energy bins for the two IXPE observations derived using the \texttt{pcube} method. }
\begin{tabular}{cccccccc}
    \hline\hline
      & \multicolumn{2}{c}{Obs.~1} & \multicolumn{2}{c}{Obs.~2} & \multicolumn{2}{c}{Obs.~1 \& 2} \\
    \hline
    Energy & $\text{PD}_\texttt{pcube}$ & $\text{PA}_\texttt{pcube}$ & $\text{PD}_\texttt{pcube}$ & $\text{PA}_\texttt{pcube}$ & $\text{PD}_\texttt{pcube}$ & $\text{PA}_\texttt{pcube}$ \\
    (keV) & (\%) & (deg) & (\%) & (deg) & (\%) & (deg) \\
    \hline
    2--3 & $10.1 \pm 3.5$ & $60 \pm 10$ & $4.5 \pm 3.4$ & $68 \pm 21$ & $7.1 \pm 2.4$ & $62 \pm 10$ \\
    3--4 & $1.3_{-1.3}^{+2.8}$ & $-43$\tablefootmark{$\dagger$} & $2.7 \pm 2.7$ & $-51 \pm 28$ & $2.0 \pm 2.0$ & $-49 \pm 28$ \\
    4--5 & $13.3 \pm 3.6$ & $74 \pm 8$ & $5.5 \pm 3.4$ & $46 \pm 18$ & $8.3 \pm 2.5$ & $66 \pm 9$ \\
    5--6 & $6.9 \pm 4.6$ & $77 \pm 19$ & $3.8_{-3.8}^{+4.5}$ & $-46$\tablefootmark{$\dagger$} & $3.1_{-3.1}^{+3.2}$ & $-86$\tablefootmark{$\dagger$} \\
    6--8 & $9.0 \pm 6.0$ & $57 \pm 19$ & $13.0 \pm 5.9$ & $45 \pm 13$ & $10.8 \pm 4.2$ & $50 \pm 11$ \\
    \hline
    2--8 & $7.2 \pm 2.3$ & $66 \pm 9$ & $3.7 \pm 2.2$ & $52 \pm 17$ & $5.2 \pm 1.6$ & $61 \pm 9$\\      
    \hline
    \end{tabular}
\tablefoot{The uncertainties are presented at the 1$\sigma$ confidence level. \tablefootmark{$(\dagger)$}{The PA cannot be confined within the 1$\sigma$ confidence level; therefore, we present the most probable value instead. }} 
\label{table:pol_ebins_pcube}
\end{table*}

The energy-resolved pulse profiles are generated using \ixpe data, with the results shown in Fig.~\ref{fig:pp-energy}. No significant evolution in the shape of the pulse profiles is observed between 3 and 6\,keV. In the softer energy band, the profiles may be influenced by absorption from the stellar wind or interstellar medium, while the harder energy band is characterized by reduced photon counts due to the instrument's lower effective area. 

\subsection{Polarimetric analysis}\label{Sect:results:pol_analysis}

The exploratory polarimetric analysis of 4U~1907+09 is performed with {\sc xspec}, fully considering the energy dispersion and the spectral shape. We first generated the Stokes $I$, $Q$, and $U$ spectra from all three DUs. We then computed the normalized Stokes parameters $q=Q/I$, $u=U/I$, the PD using the equation PD=$\sqrt{q^2+u^2}$, and the PA using PA=$\frac{1}{2} \arctan (u/q)$ measured from north to east counterclockwise on the sky \citep{Kislat_2015}. Later, we fitted all nine spectra simultaneously with {\sc xspec}. 

The spectral continuum of the source can be modeled well by an absorbed power law with a high-energy cutoff~\citep[see, e.g.,][]{Fritz_2006, Hemphill_2013, Tobrej_2023}. Given the sensitivity of IXPE in the soft X-ray bands, we adopted a simple phenomenological model: 
\begin{equation*}
\texttt{constant} \times \texttt{tbabs} \times \texttt{powerlaw} \times \texttt{polconst}\, ,
\end{equation*}
where \texttt{constant} accounts for the cross-calibration among three DUs, with the value for DU1 fixed at unity. The galactic absorption model \texttt{tbabs} characterizes the photoelectric absorption by the interstellar medium. The cross sections are provided by \citet{Verner_1996} and the element abundances are given by \citet{Wilms_2000}. We used \texttt{polconst} as the polarization model, which assumes an energy-independent PD and PA. This model is applied to both the phase-averaged and phase-resolved analysis. 
The model-independent \texttt{pcube} method was subsequently adopted as a reference for the results obtained with {\sc xspec}, as the two methods are expected to yield similar outcomes. 

\begin{table*} 
\centering
\caption{Spectro-polarimetric parameters in different pulse-phase bins for the combined dataset, derived using \textsc{xspec} and \texttt{pcube}. }
\begin{tabular}{c|ccccc|cc}
    \hline\hline
    Phase &  $N_{\mathrm{H}}$  & $\Gamma$ & ${\mathrm{PD}}_\textsc{xspec}$ & ${\mathrm{PA}}_\textsc{xspec}$ & $\chi^2$/d.o.f. & ${\mathrm{PD}}_\texttt{pcube}$ & ${\mathrm{PA}}_\texttt{pcube}$ \\ 
           & ($10^{22}\mathrm{\;cm^{-2}}$) &   & (\%) & (deg) &   & (\%) & (deg) \\
    \hline
    0.0--0.15 & $4.98_{-0.35}^{+0.36}$  & $1.37 \pm 0.08$ & $5.4 \pm 3.3$ & $-69 \pm 19$ & 689.1/752 & $7.9 \pm 4.6$ & $-86 \pm 17$ \\
    0.15--0.25 & $5.08 \pm 0.37$  & $1.04 \pm 0.08$ & $13.5 \pm 3.3$ & $71 \pm 7 $ & 825.6/758 & $17.1 \pm 4.7$ & $60 \pm 8$ \\   
    0.25--0.375 & $4.32_{-0.32}^{+0.33}$  & $1.06 \pm 0.07$ & $4.1 \pm 3.1$ & $0 \pm 24$ & 799.6/809 & $3.3_{-3.3}^{+4.3}$ & $-29$\tablefootmark{$\dagger$} \\  
    0.375--0.5 & $6.09 \pm 0.37$  & $1.36 \pm 0.08$ & ${2.4}_{-2.4}^{+3.3}$ & $73$\tablefootmark{$\dagger$} & 838.9/776 & $5.8 \pm 4.7$ & $67 \pm 23$ \\  
    0.5--0.625 & $4.74_{-0.33}^{+0.34}$  & $1.06_{-0.07}^{+0.08}$ & $4.7 \pm 3.1$ & $44 \pm 21$ & 828.1/806 & $10.4 \pm 4.4$ & $40 \pm 12$ \\   
    0.625--0.75 & $5.56 \pm 0.32$  & $1.13 \pm 0.07$ & $4.6 \pm 3.0$ & $61 \pm 20$ & 825.4/845 & $4.3 \pm 4.1$ & $73 \pm 28$ \\  
    0.750--0.85 & $4.59_{-0.35}^{+0.36}$  & $1.35 \pm 0.08$ & $4.1 \pm 3.4$ & $88 \pm 28$ & 782.3/743 & $3.5_{-3.5}^{+4.7}$ & $87$\tablefootmark{$\dagger$} \\
    0.85--1.0 & $4.95 \pm 0.35$  & $1.64_{-0.08}^{+0.09}$ & $7.1 \pm 3.4$ & $43 \pm 14$ & 695.0/728 & $8.3 \pm 4.7$ & $43 \pm 16$ \\ 
    \hline
    \end{tabular}
\tablefoot{The uncertainties in the spectro-polarimetric parameters, computed using the \texttt{error} command in \textsc{xspec}, are provided at the 68.3\% (1$\sigma$) confidence level ($\Delta\chi^2 = 1$ for one parameter of interest). 
The right two columns give PDs and PAs estimated with the \texttt{pcube} algorithm and the uncertainties given at the 1$\sigma$ confidence level. \tablefootmark{$(\dagger)$}{The PA cannot be confined within the 1$\sigma$ confidence level; therefore, we presented the most probable value instead. } } 
\label{table:phase-res-xspec-tot}
\end{table*}

\subsubsection{Phase-averaged analysis} \label{Sect:results:pol_analysis:phase-avg} 

We began with a phase-averaged analysis in the full 2--8\,keV band for both observations, with the results summarized in Table~\ref{table:phase_avg}. Obs.~1 exhibits stronger polarization than Obs.~2, with phase-averaged PD of $6.0 \pm 1.6\%$ and $2.2 \pm 1.6\%$, respectively. The corresponding PAs are $69\degr \pm 8\degr$ for Obs.~1 and $46\degr \pm 23\degr$ for Obs.~2. We also performed a spectro-polarimetric analysis on the combined dataset from the two observations, with the results summarized in the last column of Table~\ref{table:phase_avg}. Assuming a source distance of $d \simeq 1.9$\,kpc~\citep{Bailer-Jones_2021}, we estimate the luminosity of 4U~1907+09 to be $L \simeq 2.5 \times {10}^{35}$\,erg\,s$^{-1}$.

\begin{figure}
\centering
\includegraphics[width=0.95\linewidth]{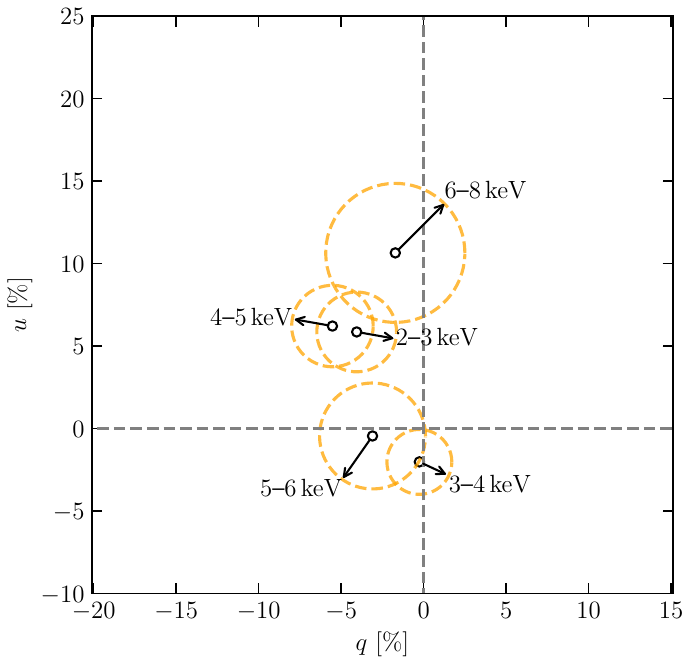}
\caption{Energy dependence of the normalized Stokes parameters $q$ and $u$ for the phase-averaged data of the combined dataset (Obs.~1 and 2), obtained with the \texttt{pcube} algorithm. The 1$\sigma$ contours are plotted as dashed circles around each best estimate. }
\label{fig:QU-energy}
\end{figure}

Next, we studied the energy dependence of the polarization properties of 4U~1907+09 by performing an energy-resolved polarimetric analysis on each observation, dividing the data into five energy bins (2--3, 3--4, 4--5, 5--6, and 6--8\,keV). We performed the fitting with {\sc xspec}, using all energy bins within 2--8\,keV for the Stokes $I$ spectra while excluding bins outside the interested energy range for the Stokes $Q$ and $U$ spectra. The results are shown in Table~\ref{table:pol_ebins_xspec}. We also conducted the polarimetric analysis with the model-independent \texttt{pcube} algorithm, with the results shown in Table~\ref{table:pol_ebins_pcube}, demonstrating similar outcomes.

Contrary to the expectation that the PA is independent of energy, we observed a larger PD with greater confidence in the 2--3, 4--5, and 6--8\,keV bands during the two \ixpe observations. Furthermore, polarimetric detections in different energy bands exhibit distinct PAs. The normalized Stokes parameters $q$ and $u$ for each energy band of the combined dataset, as shown in Fig.~\ref{fig:QU-energy}, further highlight the polarization discrepancies across energy bands. 

The energy dependence of the PA is also observed in other XRPs. In Vela~X-1, the PA differs by approximately 90\degr\ between the 2--3 and 3--8\,keV bands \citep{Forsblom_2023_VelaX1, Forsblom_2025_VelaX1}. Similarly, in 4U~1538$-$52, the PA exhibits a shift of about 70\degr\ between the 2--3 and 4--8~keV bands \citep{Loktev_2025_4U1538}.

\begin{figure*}
\centering
\includegraphics[width=0.33\textwidth]{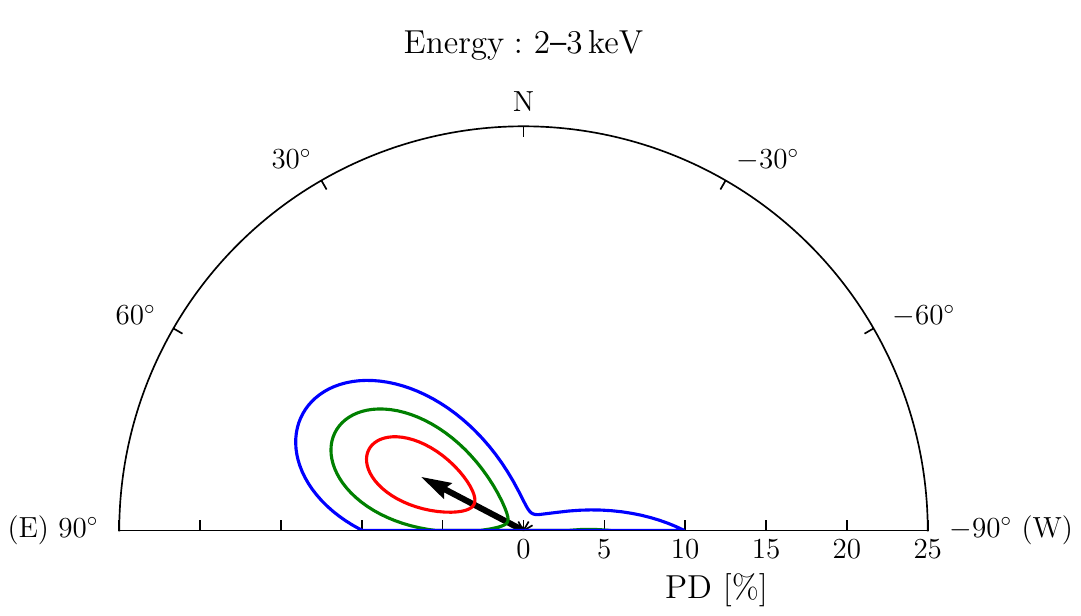}
\includegraphics[width=0.33\textwidth]{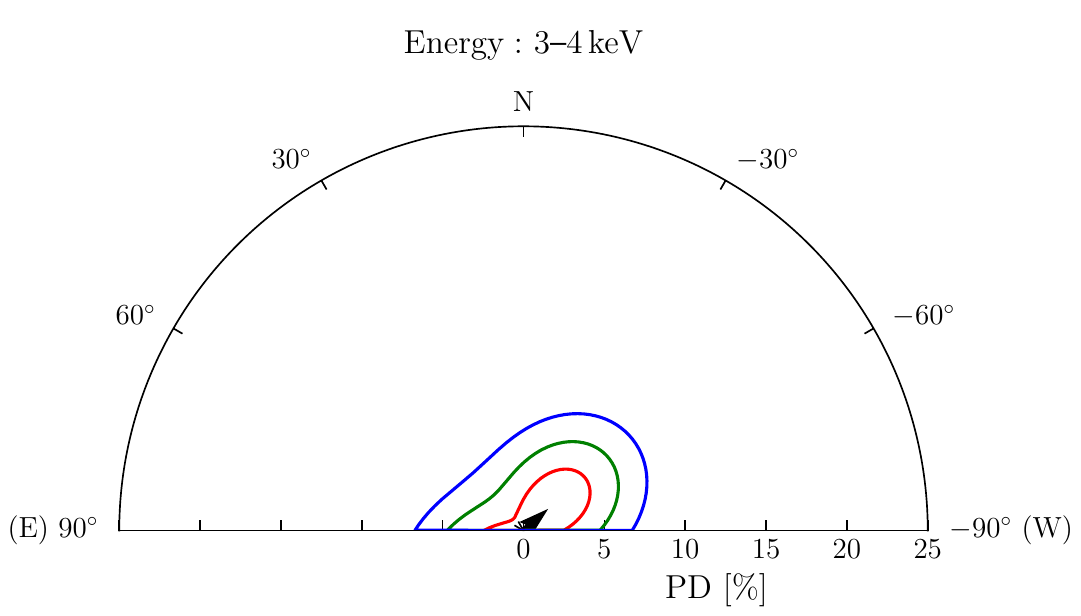}
\includegraphics[width=0.33\textwidth]{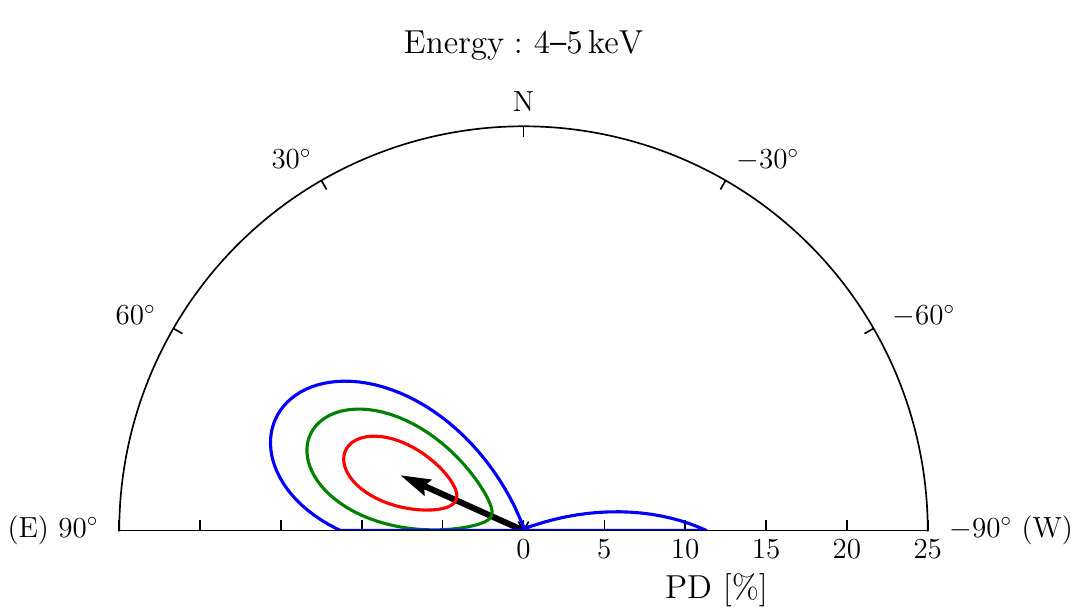}
\includegraphics[width=0.33\textwidth]{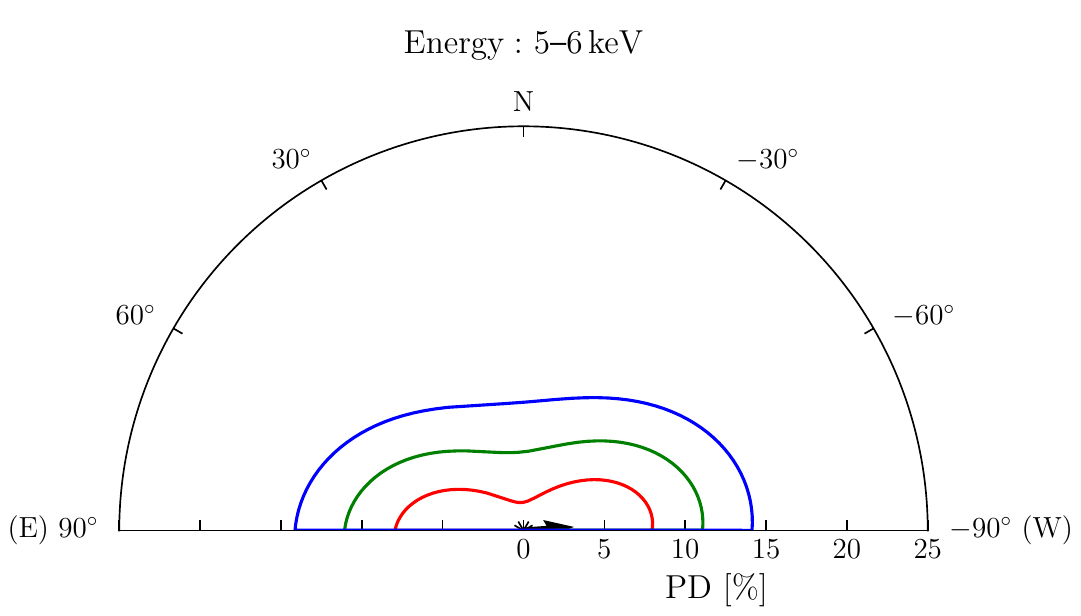}
\includegraphics[width=0.33\textwidth]{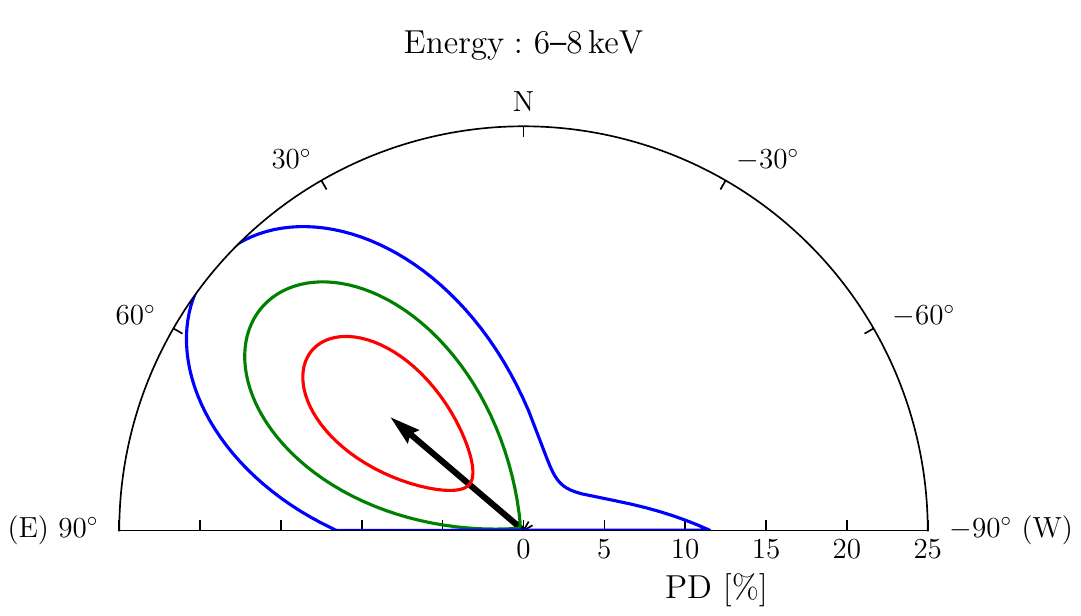}
\caption{Polarization vectors of 4U~1907+09 across different energy bands derived from the phase-averaged polarimetric analysis of the combined dataset using the \texttt{pcube} algorithm. Confidence contours at 68.3\%, 95.45\%, and 99.73\% are displayed in red, green, and blue, respectively.}
\label{fig:protractors-energy}
\end{figure*}

We then tested the null hypothesis that the PA does not depend on energy. Since PAs are not normally distributed, we adopted the probability density function of the PA, $\psi$, as derived by \citet{Naghizadeh-Khouei_1993}:
\begin{equation} \label{eq:PA_dist}
G(\psi) = \frac{1}{\sqrt{\pi}}
\left\{  \frac{1}{\sqrt{\pi}}  + 
\eta {\rm e}^{\eta^2} 
\left[ 1 + {\rm erf}(\eta) \right]
\right\} {\rm e}^{-p_0^2/2} \, , \notag
\end{equation}
where $p_0 = \text{PD}/\text{PD}_\text{err}$ is the measured PD in units of its error, $\eta = p_0 \cos\left[ 2 \left( \psi-\psi_0 \right) \right]/\sqrt{2}$, and $\psi_0$ is the expectation value of the PA. Here, \mbox{erf} denotes the error function. To evaluate the statistical significance, we employed the parameters obtained from the {\sc xspec} fit (see Table~\ref{table:pol_ebins_xspec}) to perform Monte Carlo simulations. We defined the statistical test quantity, the $Z$-score, as $Z = - \sum \log G (\psi)$. To simulate the probability distribution of the $Z$-score under the null hypothesis, we used the average PA ($63\degr$), PD (3.7\%) and its uncertainty (1.1\%) within the 2--8\,keV range. The observed $Z$-score is calculated using the most probable values of $\psi$ of the individual energy bands, with $p_0$ determined based on the corresponding PDs and their positive uncertainties, and setting $\psi_0 = 63\degr$. Our analysis yields a $p$-value of 0.049 to reject the null hypothesis, corresponding to a statistical significance of 1.7\,$\sigma$. This result suggests tentative evidence that the PA depends on the energy. 

\begin{figure}
\centering
\includegraphics[width=0.9\linewidth]{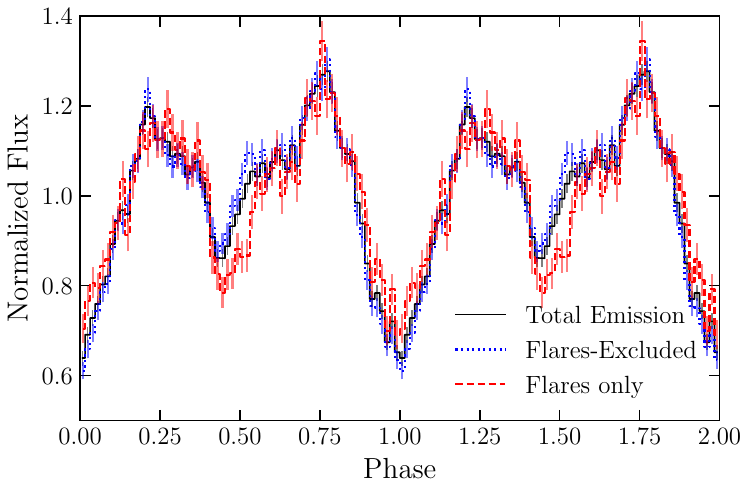}
\caption{Pulse profiles of 4U~1907+09 from the combined \ixpe observation. Dotted blue, dashed red, and solid black lines represent the profiles for flares-excluded data, flare-only data, and total emission, respectively. The pulse profile fluxes are normalized such that their mean values are unity. }
\label{fig:pp-flares}
\end{figure}

\subsubsection{Phase-resolved analysis} \label{Sect:results:pol_analysis:phase-res} 

Studying the phase-dependent properties of pulsars is crucial, as their geometries change with rotation. Applying the ephemerides in Table~\ref{table:time-pars}, the phase of each photon event can be determined by 
\begin{equation*}
\phi = \frac{1}{P} (t - t_0) - \frac{1}{2} \frac{\dot{P}}{P^2} ( t - t_0 )^2 \, ,
\end{equation*}
in which $P$ is the pulse period and $\dot{P}$ is the pulse period derivative at a reference time $t_0$, which also defines the zero phase. The photons are divided into eight phase bins, selected to ensure that each bin contains approximately the same number of photon counts. For each phase bin, we generated the Stokes $I$, $Q$, and $U$ spectra, as described in Sect.~\ref{Sect:data}. 

We first applied the {\sc xspec} method to each phase bin, focusing on the 2--8\,keV energy range. The results of the combined dataset are presented in Table~\ref{table:phase-res-xspec-tot}. Although a few phase bins show polarization detections with a significance level below 1$\sigma$, the modulation of the PA with pulse phase remains detectable despite the gaps in phase coverage. 

The \texttt{pcube} algorithm is subsequently applied to the phase-resolved data, with the results shown in the last two columns of Table~\ref{table:phase-res-xspec-tot}. The PDs obtained using the \texttt{pcube} method are generally higher than those derived with \textsc{xspec}, while the PAs from both methods are typically well aligned when the respective PDs are significant. 

\subsection{Flares and their impacts on polarimetric properties}\label{Sect:results:flares}
To date, the impact of flares on polarimetric detection remains unclear. To investigate this influence, we analyzed the events detected during the flares separately from those detected outside of the flares. 

We first examined the pulse profiles during the flare and non-flare periods. The results, along with a comparison to the pulse profile of the total emission, are shown in Fig.~\ref{fig:pp-flares}. The pulse profiles exhibit subtle differences, suggesting that the geometry of the accreting pulsar undergoes slight changes during flares. 

\begin{figure*}
\centering
\includegraphics[width=0.85\linewidth]{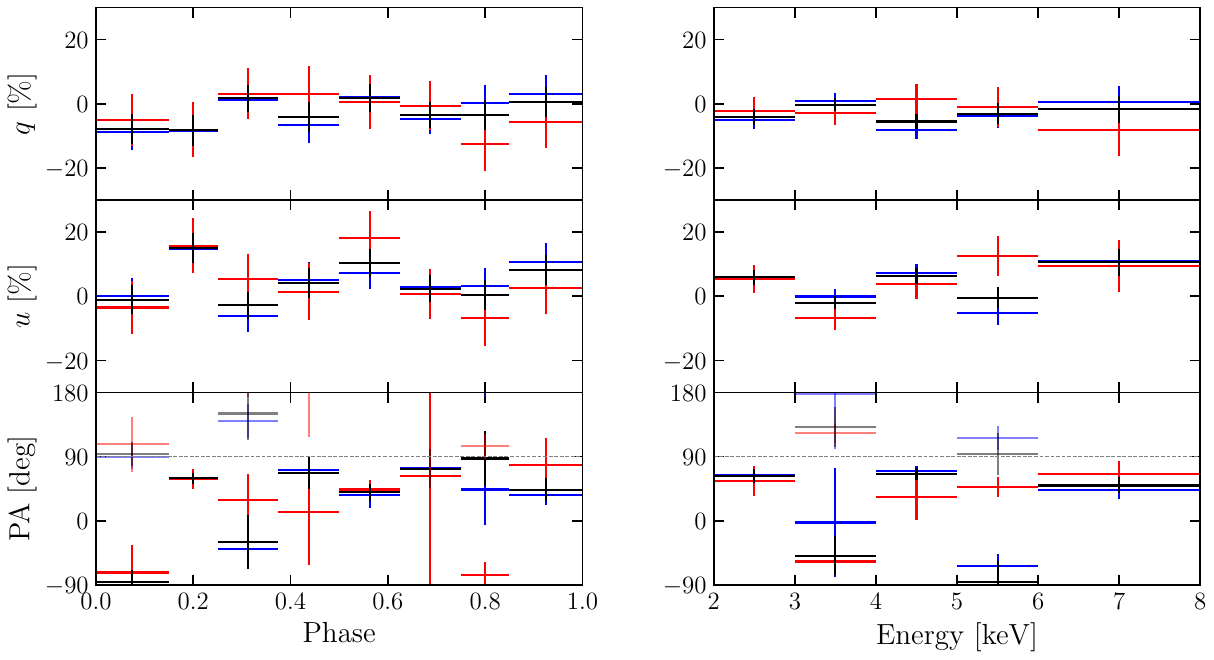}
\caption{Normalized Stokes parameters and measured PAs as functions of pulse phase and energy from the combined \ixpe observation. Blue, red, and black crosses represent the flares-excluded data, flare-only data, and total emission data, respectively. Some PAs are plotted twice with an added constant of 180\degr\ to illustrate the continuity of the PA with respect to phase and energy.}
\label{fig:QU-flares}
\end{figure*}

We then investigated the behavior of the Stokes parameters as functions of pulse phase and energy. The normalized Stokes parameters $q$, $u$, and the derived PAs across different phase and energy intervals are presented in Fig.~\ref{fig:QU-flares}. The uncertainties of these parameters during flares are larger due to the lower photon counts in the flare periods. However, the parameters are consistent within their respective uncertainties. 

\begin{figure}
\centering
\includegraphics[width=0.49\textwidth]{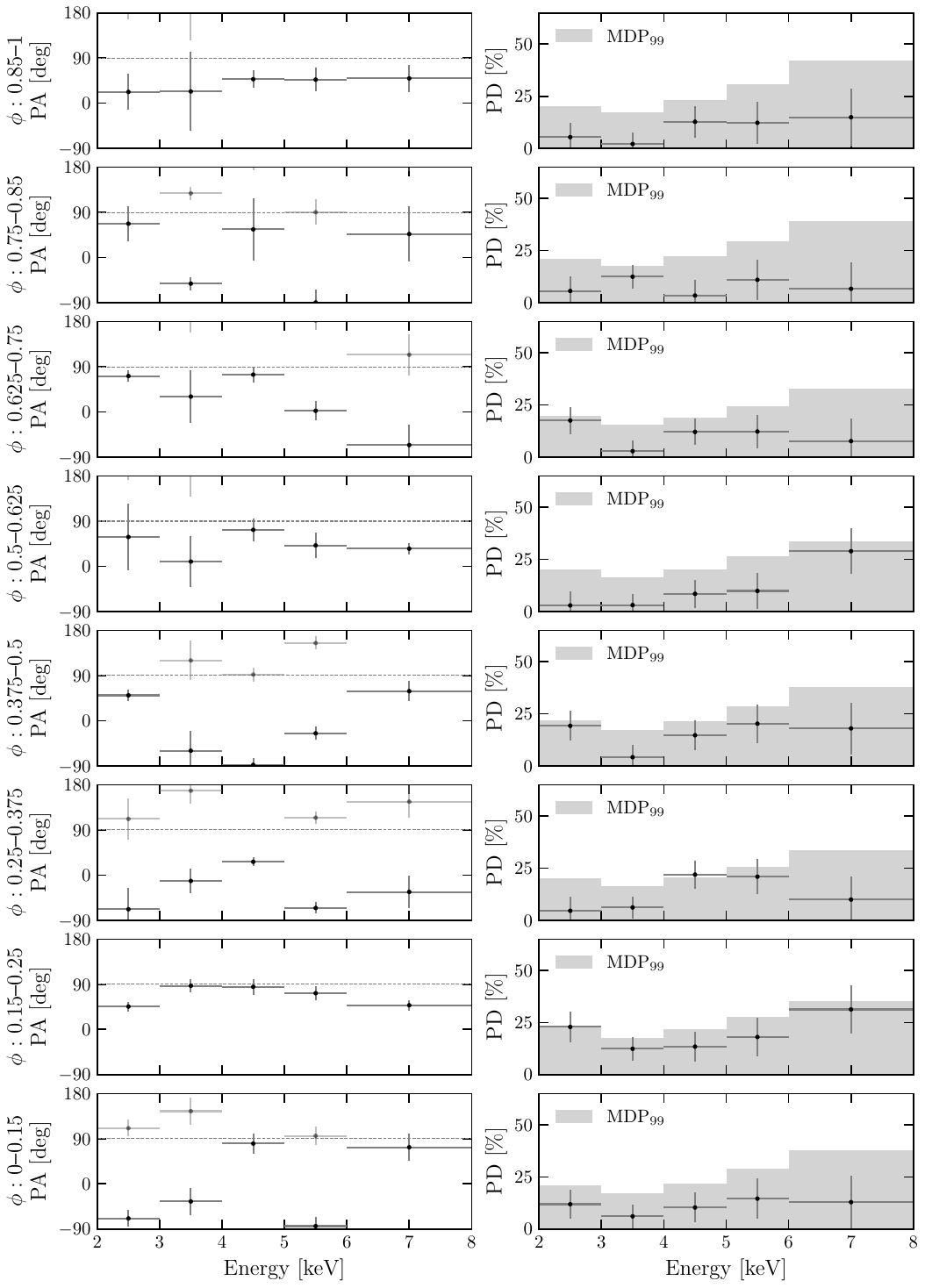}
\caption{Energy-resolved polarization measurements of the combined dataset across pulse phases. Left: PAs with 1$\sigma$ uncertainties. Right: PDs. Gray bars indicate the $\mathrm{MDP}_{99}$ value. Note that we again plot the PAs with an added constant of 180$\degr$ to illustrate the continuity of the PAs versus energy. }
\label{fig:phase_energy_PA}
\end{figure}

Given that 4U~1907+09 is not particularly bright and the photon counts during flares are insufficient, we conclude that the energy-dependent polarimetric properties are not a result of the flares. In fact, flares have minimal impact on the observed polarimetric properties. Therefore, in the following analysis, we include all emissions from the source without distinguishing between flares and non-flares. 

\subsection{Energy-dependent polarimetric properties in different pulse phases}\label{Sect:results:energy-phase-resolved}

We have observed energy-dependent polarimetric properties in the phase-averaged analysis, and in Sect.~\ref{Sect:results:flares} we concluded that they are unlikely to be produced by flares. It is then necessary to examine the energy-dependent polarimetric properties in each pulse phase. Accordingly, we divided the phase-tagged photon events into eight phase bins, as in Table~\ref{table:phase-res-xspec-tot}, and into five energy bins covering the 2--8\,keV range, following Tables~\ref{table:pol_ebins_xspec} and \ref{table:pol_ebins_pcube}. The \texttt{pcube} algorithm was then used to compute the PA and PD in each energy-phase bin. The results are shown in Fig.~\ref{fig:phase_energy_PA}. 

\begin{figure}
\centering
\includegraphics[width=0.85\linewidth]{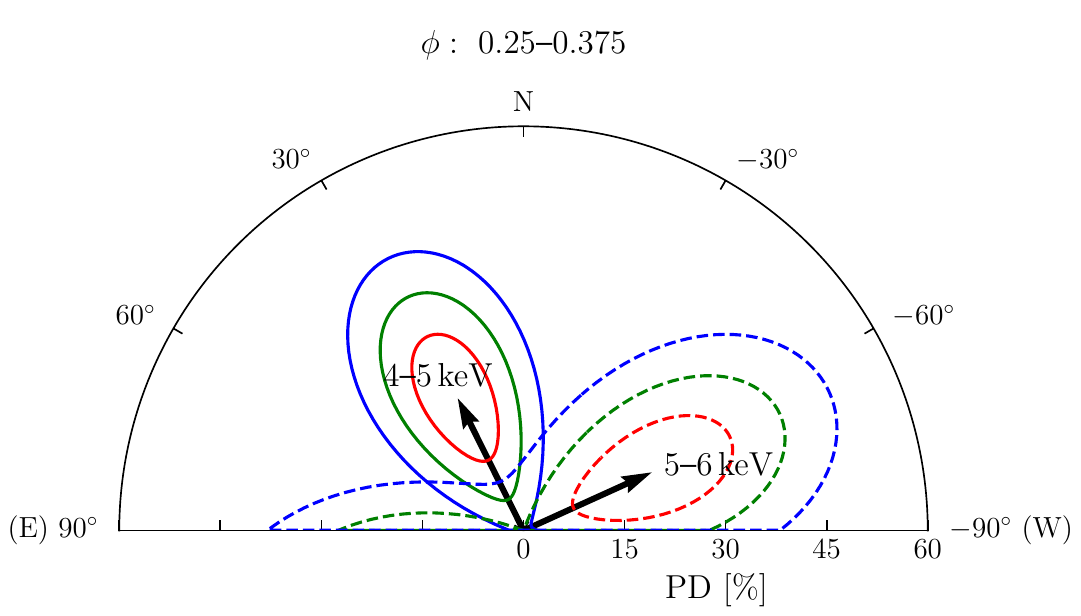}
\caption{Protractor plot of the polarization measurement of the combined dataset within the 4--5\,keV band and the 5--6\,keV band in the pulse phase interval 0.25--0.375. }
\label{fig:protractor}
\end{figure}

Due to the limited photon counts, most PDs in the energy-phase bins do not exceed the minimum detectable polarization at the 99\% confidence level ($\mathrm{MDP}_{99}$). Nonetheless, the evolution of the PA with energy is still observable in each phase bin. 
The most notable detection of a PA rotation between adjacent energy bands within a single phase bin occurs in the phase interval 0.25--0.375. In this interval, the PA of photons in the 4--5\,keV and 5--6\,keV energy bands differs by approximately 90\degr. Their PDs marginally reach the $\mathrm{MDP}_{99}$, suggesting a ``probable'' right-angle rotation of the PA between these energy bands within this phase interval. Protractor plots illustrating the PAs with their 1$\sigma$, 2$\sigma$, and 3$\sigma$ uncertainties are shown in Fig.~\ref{fig:protractor}, where the PAs are compatible only within the 3$\sigma$ range. It is worth noting that in Vela~X-1, a PA shift of $\sim 90\degr$ between two energy bands is also detected within a single phase bin \citep{Forsblom_2025_VelaX1}. 

\section{Discussion}\label{Sect:discussion}

X-ray Pulsars are considered prime candidates for X-ray polarimetric analysis, as the highly magnetized matter accreting onto the magnetic poles of the NS can naturally generate a high PD due to the birefringence of the medium in such a strong magnetic field. The oscillations of the electric field of the emitted photons are expected to align either parallel or perpendicular to the plane formed by the photon momentum and the local magnetic field \citep{Gnedin_1974, Gnedin_1978}. These modes are referred to as the ordinary (O) and extraordinary (X), respectively. The Compton scattering cross section below the cyclotron energy for the O-mode radiation is much higher than that for the X-mode \citep{Lai_2003}, predicting a high PD up to 80\% \citep{Meszaros_1988_RVM, Caiazzo_2021}. 

Nevertheless, most XRPs observed by \ixpe exhibit low phase-averaged PDs \citep[see, e.g.,][]{Tsygankov_2022_CenX3, Mushtukov_2023_XPersei, Forsblom_2023_VelaX1}, with higher PDs emerging in phase-resolved analyses. This trend does not fully hold for 4U~1907+09, where polarization is weakly detected in most phase bins. Further examination of the energy dependence of the PA within individual phase bins partially enhances the significance of polarization detection in some energy-phase bins, as shown in Fig.~\ref{fig:phase_energy_PA}. However, some energy-phase bins still exhibit extremely low PD significance. 

Because the measured PAs exhibit energy dependence in the phase-averaged analysis, the rotating vector model (RVM; \citealt{Radhakrishnan_1969_RVM, Meszaros_1988_RVM, Poutanen_2020_RVM}), which assumes that the PA aligns with the local magnetic field and is therefore energy-independent, is not suitable for constraining the pulsar's geometry. Furthermore, the absence of a distinct PA transition at any specific energy further precludes the application of an energy-resolved RVM fit \citep[e.g.,][]{Forsblom_2025_VelaX1, Loktev_2025_4U1538}. 

Various accretion processes involving reflection and scattering can introduce additional polarized components that tangle with the emission produced near the magnetic poles \citep[see the discussion in][]{Tsygankov_2022_CenX3}. Consequently, it is natural to consider an energy-dependent polarized component as a potential explanation for the unusual energy dependence of the PA. This idea has been explored by \citet{Forsblom_2025_VelaX1} and \citet{Loktev_2025_4U1538}. Specifically, \citet{Forsblom_2025_VelaX1} propose two possible mechanisms for the observed PA variation with energy. The first involves a spectral component that diminishes at higher energies and generally has a PA orthogonal to that of the dominant high-energy component. The second invokes vacuum resonance, in which one polarization mode converts into another near the cyclotron line energy, resulting in a $90\degr$ shift in the PA. Given the results shown in Tables~\ref{table:pol_ebins_xspec} and \ref{table:pol_ebins_pcube}, it is difficult to envision a spectral component that is extraordinarily strong specifically in the 3--4 and 5--6\,keV bands, thereby causing the observed drop in PDs. Regarding the second mechanism, it predicts an energy-independent PA at both low and high energies. However, this prediction cannot be confirmed due to the limited photon statistics ($\simeq 1.45$ million photons within the 2--8\,keV band for the combined dataset) and the restricted energy coverage of \ixpe. Furthermore, if a gradual shift in the PA with energy is observed, it would indicate the presence of an additional polarized component. More robust results are expected from future observations with improved photon statistics and broader energy coverage. 

\section{Summary}\label{Sect:summary}

4U~1907+09 was observed by \ixpe twice during the period 2024 November 11--21, during its brightest periods close to the periastron. We summarize the results of the polarimetric analysis of 4U~1907+09 as follows: 
\begin{enumerate}
\item A more significant polarization is detected in Obs.~1 than in Obs.~2, with phase-averaged PD and PA values of $6.0 \pm 1.6\%$ and $69\degr \pm 8\degr$ for Obs.~1 and $2.2 \pm 1.6\%$ and $46\degr \pm 23\degr$ for Obs.~2. The combined dataset from the two observations yields a PD of $3.7 \pm 1.1\%$ and a PA of $63\degr \pm 9\degr$. 
\item An energy-dependent PA is likely observed in the phase-averaged polarimetric analyses with a significance of 1.7\,$\sigma$, partially explaining the generally insignificant polarization detections. Specifically, we detect a probable rotation of the PA between adjacent energy bands within one single phase bin. 
\item We further examined whether the three brief flares detected by \ixpe significantly affect the pulsar’s polarization. Our analysis concludes that the polarimetric properties are similar in the flare-excluded and flare-only data, suggesting that the observed energy-dependent PAs are unlikely due to the flares. 
\end{enumerate}

\begin{acknowledgements}
The Imaging X-ray Polarimetry Explorer (\ixpe) is a joint US and Italian mission. The US contribution is supported by the National Aeronautics and Space Administration (NASA) and led and managed by its Marshall Space Flight Center (MSFC), with industry partner Ball Aerospace (contract NNM15AA18C). The Italian contribution is supported by the Italian Space Agency (Agenzia Spaziale Italiana, ASI) through contract ASI-OHBI-2022-13-I.0, agreements ASI-INAF-2022-19-HH.0 and ASI-INFN-2017.13-H0, and its Space Science Data Center (SSDC) with agreements ASI-INAF-2022-14-HH.0 and ASI-INFN 2021-43-HH.0, and by the Istituto Nazionale di Astrofisica (INAF) and the Istituto Nazionale di Fisica Nucleare (INFN) in Italy. This research used data products provided by the IXPE Team (MSFC, SSDC, INAF, and INFN) and distributed with additional software tools by the High-Energy Astrophysics Science Archive Research Center (HEASARC), at NASA Goddard Space Flight Center (GSFC). 
This research has been supported by the Deutsche Forschungsgemeinschaft (DFG) grants 549824807 (LD) and WE 1312/59-1 (VFS), Vilho, Yrjö, and Kalle Väisälä foundation (SVF), and the UKRI Stephen Hawking fellowship (AAM). 
ASal acknowledges funding from the EDUFI Fellowship and Jenny and Antti Wihuri Foundation.
ADM and FLM are partially supported by MAECI with grant CN24GR08 ``GRBAXP: Guangxi-Rome Bilateral Agreement for X-ray Polarimetry in Astrophysics''.
\end{acknowledgements}

\bibliographystyle{yahapj}
\bibliography{references}

\begin{thebibliography}{}
\providecommand\natexlab[1]{#1}
\providecommand\JournalTitle[1]{#1}

\bibitem[{{Arnaud}(1996)}]{Arnaud_1996}
{Arnaud}, K.~A. 1996, in ASP Conf. Ser., Vol. 101, Astronomical Data Analysis Software and Systems V, ed. G.~H. {Jacoby} \& J.~{Barnes} (San Francisco: Astron. Soc. Pac.), 17

\bibitem[{{Bailer-Jones} {et~al.}(2021){Bailer-Jones}, {Rybizki}, {Fouesneau}, {Demleitner}, \& {Andrae}}]{Bailer-Jones_2021}
{Bailer-Jones}, C.~A.~L., {Rybizki}, J., {Fouesneau}, M., {Demleitner}, M., \& {Andrae}, R. 2021, \href{http://dx.doi.org/10.3847/1538-3881/abd806}{\JournalTitle{\aj}, 161, 147}

\bibitem[{{Baldini} {et~al.}(2021){Baldini}, {Barbanera}, {Bellazzini}, {Bonino}, {Borotto}, {Brez}, {Caporale}, {Cardelli}, {Castellano}, {Ceccanti}, {Citraro}, {Di Lalla}, {Latronico}, {Lucchesi}, {Magazz{\`u}}, {Magazz{\`u}}, {Maldera}, {Manfreda}, {Marengo}, {Marrocchesi}, {Mereu}, {Minuti}, {Mosti}, {Nasimi}, {Nuti}, {Oppedisano}, {Orsini}, {Pesce-Rollins}, {Pinchera}, {Profeti}, {Sgr{\`o}}, {Spandre}, {Tardiola}, {Zanetti}, {Amici}, {Andersson}, {Attin{\`a}}, {Bachetti}, {Baumgartner}, {Brienza}, {Carpentiero}, {Castronuovo}, {Cavalli}, {Cavazzuti}, {Centrone}, {Costa}, {D'Alba}, {D'Amico}, {Del Monte}, {Di Cosimo}, {Di Marco}, {Di Persio}, {Donnarumma}, {Evangelista}, {Fabiani}, {Ferrazzoli}, {Kitaguchi}, {La Monaca}, {Lefevre}, {Loffredo}, {Lorenzi}, {Mangraviti}, {Matt}, {Meilahti}, {Morbidini}, {Muleri}, {Nakano}, {Negri}, {Nenonen}, {O'Dell}, {Perri}, {Piazzolla}, {Pieraccini}, {Pilia}, {Puccetti}, {Ramsey}, {Rankin}, {Ratheesh}, {Rubini}, {Santoli}, {Sarra}, {Scalise}, {Sciortino}, {Soffitta},
  {Tamagawa}, {Tennant}, {Tobia}, {Trois}, {Uchiyama}, {Vimercati}, {Weisskopf}, {Xie}, {Zanetti}, \& {Zhou}}]{Baldini_2021}
{Baldini}, L., {Barbanera}, M., {Bellazzini}, R., {et~al.} 2021, \href{http://dx.doi.org/10.1016/j.astropartphys.2021.102628}{\JournalTitle{Astroparticle Physics}, 133, 102628}

\bibitem[{{Baldini} {et~al.}(2022){Baldini}, {Bucciantini}, {Lalla}, {Ehlert}, {Manfreda}, {Negro}, {Omodei}, {Pesce-Rollins}, {Sgr{\`o}}, \& {Silvestri}}]{Baldini_2022_ixpeobssim}
{Baldini}, L., {Bucciantini}, N., {Lalla}, N.~D., {et~al.} 2022, \href{http://dx.doi.org/10.1016/j.softx.2022.101194}{\JournalTitle{SoftwareX}, 19, 101194}

\bibitem[{{Buccheri} {et~al.}(1983){Buccheri}, {Bennett}, {Bignami}, {Bloemen}, {Boriakoff}, {Caraveo}, {Hermsen}, {Kanbach}, {Manchester}, {Masnou}, {Mayer-Hasselwander}, {{\"O}zel}, {Paul}, {Sacco}, {Scarsi}, \& {Strong}}]{Buccheri_1983}
{Buccheri}, R., {Bennett}, K., {Bignami}, G.~F., {et~al.} 1983, \JournalTitle{\aap}, 128, 245

\bibitem[{{Caiazzo} \& {Heyl}(2021)}]{Caiazzo_2021}
{Caiazzo}, I., \& {Heyl}, J. 2021, \href{http://dx.doi.org/10.1093/mnras/staa3428}{\JournalTitle{\mnras}, 501, 109}

\bibitem[{{Cox} {et~al.}(2005){Cox}, {Kaper}, \& {Mokiem}}]{Cox_2005}
{Cox}, N.~L.~J., {Kaper}, L., \& {Mokiem}, M.~R. 2005, \href{http://dx.doi.org/10.1051/0004-6361:20040511}{\JournalTitle{\aap}, 436, 661}

\bibitem[{{{\c{S}}ahiner} {et~al.}(2012){{\c{S}}ahiner}, {Inam}, \& {Baykal}}]{Sahiner_2012}
{{\c{S}}ahiner}, {\c{S}}., {Inam}, S.~{\c{C}}., \& {Baykal}, A. 2012, \href{http://dx.doi.org/10.1111/j.1365-2966.2012.20455.x}{\JournalTitle{\mnras}, 421, 2079}

\bibitem[{{Cusumano} {et~al.}(2000){Cusumano}, {di Salvo}, {Burderi}, {Orlandini}, {Piraino}, {Robba}, \& {Santangelo}}]{Cusumano_2000}
{Cusumano}, G., {di Salvo}, T., {Burderi}, L., {et~al.} 2000, \href{http://dx.doi.org/10.1016/S0273-1177(99)00768-1}{\JournalTitle{Advances in Space Research}, 25, 409}

\bibitem[{{Di Marco} {et~al.}(2022){Di Marco}, {Costa}, {Muleri}, {Soffitta}, {Fabiani}, {La Monaca}, {Rankin}, {Xie}, {Bachetti}, {Baldini}, {Baumgartner}, {Bellazzini}, {Brez}, {Castellano}, {Del Monte}, {Di Lalla}, {Ferrazzoli}, {Latronico}, {Maldera}, {Manfreda}, {O'Dell}, {Perri}, {Pesce-Rollins}, {Puccetti}, {Ramsey}, {Ratheesh}, {Sgr{\`o}}, {Spandre}, {Tennant}, {Tobia}, {Trois}, \& {Weisskopf}}]{DiMarco_2022}
{Di Marco}, A., {Costa}, E., {Muleri}, F., {et~al.} 2022, \href{http://dx.doi.org/10.3847/1538-3881/ac51c9}{\JournalTitle{\aj}, 163, 170}

\bibitem[{{Di Marco} {et~al.}(2023){Di Marco}, {Soffitta}, {Costa}, {Ferrazzoli}, {La Monaca}, {Rankin}, {Ratheesh}, {Xie}, {Baldini}, {Del Monte}, {Ehlert}, {Fabiani}, {Kim}, {Muleri}, {O'Dell}, {Ramsey}, {Rubini}, {Sgr{\`o}}, {Silvestri}, {Tennant}, \& {Weisskopf}}]{DiMarco_2023}
{Di Marco}, A., {Soffitta}, P., {Costa}, E., {et~al.} 2023, \href{http://dx.doi.org/10.3847/1538-3881/acba0f}{\JournalTitle{\aj}, 165, 143}

\bibitem[{{Elsner} \& {Lamb}(1977)}]{Elsner_1977}
{Elsner}, R.~F., \& {Lamb}, F.~K. 1977, \href{http://dx.doi.org/10.1086/155427}{\JournalTitle{\apj}, 215, 897}

\bibitem[{{Ferrigno} {et~al.}(2022){Ferrigno}, {Bozzo}, \& {Romano}}]{Ferrigno_2022}
{Ferrigno}, C., {Bozzo}, E., \& {Romano}, P. 2022, \href{http://dx.doi.org/10.1051/0004-6361/202243294}{\JournalTitle{\aap}, 664, A99}

\bibitem[{{Forman} {et~al.}(1978){Forman}, {Jones}, {Cominsky}, {Julien}, {Murray}, {Peters}, {Tananbaum}, \& {Giacconi}}]{Forman_1978}
{Forman}, W., {Jones}, C., {Cominsky}, L., {et~al.} 1978, \href{http://dx.doi.org/10.1086/190561}{\JournalTitle{\apjs}, 38, 357}

\bibitem[{{Forsblom} {et~al.}(2025){Forsblom}, {Tsygankov}, {Suleimanov}, {Mushtukov}, \& {Poutanen}}]{Forsblom_2025_VelaX1}
{Forsblom}, S.~V., {Tsygankov}, S.~S., {Suleimanov}, V.~F., {Mushtukov}, A.~A., \& {Poutanen}, J. 2025, \href{http://dx.doi.org/10.1051/0004-6361/202553867}{\JournalTitle{\aap}, 696, A224}

\bibitem[{{Forsblom} {et~al.}(2023){Forsblom}, {Poutanen}, {Tsygankov}, {Bachetti}, {Di Marco}, {Doroshenko}, {Heyl}, {La Monaca}, {Malacaria}, {Marshall}, {Muleri}, {Mushtukov}, {Pilia}, {Rogantini}, {Suleimanov}, {Taverna}, {Xie}, {Agudo}, {Antonelli}, {Baldini}, {Baumgartner}, {Bellazzini}, {Bianchi}, {Bongiorno}, {Bonino}, {Brez}, {Bucciantini}, {Capitanio}, {Castellano}, {Cavazzuti}, {Chen}, {Ciprini}, {Costa}, {De Rosa}, {Del Monte}, {Di Gesu}, {Di Lalla}, {Donnarumma}, {Dov{\v{c}}iak}, {Ehlert}, {Enoto}, {Evangelista}, {Fabiani}, {Ferrazzoli}, {Garcia}, {Gunji}, {Hayashida}, {Iwakiri}, {Jorstad}, {Kaaret}, {Karas}, {Kitaguchi}, {Kolodziejczak}, {Krawczynski}, {Latronico}, {Liodakis}, {Maldera}, {Manfreda}, {Marin}, {Marinucci}, {Marscher}, {Matt}, {Mitsuishi}, {Mizuno}, {Negro}, {Ng}, {O'Dell}, {Omodei}, {Oppedisano}, {Papitto}, {Pavlov}, {Peirson}, {Perri}, {Pesce-Rollins}, {Petrucci}, {Possenti}, {Puccetti}, {Ramsey}, {Rankin}, {Ratheesh}, {Roberts}, {Romani}, {Sgr{\`o}}, {Slane}, {Soffitta},
  {Spandre}, {Sunyaev}, {Swartz}, {Tamagawa}, {Tavecchio}, {Tawara}, {Tennant}, {Thomas}, {Tombesi}, {Trois}, {Turolla}, {Vink}, {Weisskopf}, {Wu}, {Zane}, \& {IXPE Collaboration}}]{Forsblom_2023_VelaX1}
{Forsblom}, S.~V., {Poutanen}, J., {Tsygankov}, S.~S., {et~al.} 2023, \href{http://dx.doi.org/10.3847/2041-8213/acc391}{\JournalTitle{\apjl}, 947, L20}

\bibitem[{{Fritz} {et~al.}(2006){Fritz}, {Kreykenbohm}, {Wilms}, {Staubert}, {Bayazit}, {Pottschmidt}, {Rodriguez}, \& {Santangelo}}]{Fritz_2006}
{Fritz}, S., {Kreykenbohm}, I., {Wilms}, J., {et~al.} 2006, \href{http://dx.doi.org/10.1051/0004-6361:20065557}{\JournalTitle{\aap}, 458, 885}

\bibitem[{{F{\"u}rst} {et~al.}(2012){F{\"u}rst}, {Pottschmidt}, {Kreykenbohm}, {M{\"u}ller}, {K{\"u}hnel}, {Wilms}, \& {Rothschild}}]{Fuerst_2012}
{F{\"u}rst}, F., {Pottschmidt}, K., {Kreykenbohm}, I., {et~al.} 2012, \href{http://dx.doi.org/10.1051/0004-6361/201219845}{\JournalTitle{\aap}, 547, A2}

\bibitem[{{Giacconi} {et~al.}(1971){Giacconi}, {Kellogg}, {Gorenstein}, {Gursky}, \& {Tananbaum}}]{Giacconi_1971}
{Giacconi}, R., {Kellogg}, E., {Gorenstein}, P., {Gursky}, H., \& {Tananbaum}, H. 1971, \href{http://dx.doi.org/10.1086/180711}{\JournalTitle{\apjl}, 165, L27}

\bibitem[{{Gnedin} \& {Pavlov}(1974)}]{Gnedin_1974}
{Gnedin}, Y.~N., \& {Pavlov}, G.~G. 1974, \JournalTitle{Soviet Journal of Experimental and Theoretical Physics}, 38, 903

\bibitem[{{Gnedin} {et~al.}(1978){Gnedin}, {Pavlov}, \& {Shibanov}}]{Gnedin_1978}
{Gnedin}, Y.~N., {Pavlov}, G.~G., \& {Shibanov}, Y.~A. 1978, \JournalTitle{Soviet Astronomy Letters}, 4, 117

\bibitem[{{Hemphill} {et~al.}(2013){Hemphill}, {Rothschild}, {Caballero}, {Pottschmidt}, {K{\"u}hnel}, {F{\"u}rst}, \& {Wilms}}]{Hemphill_2013}
{Hemphill}, P.~B., {Rothschild}, R.~E., {Caballero}, I., {et~al.} 2013, \href{http://dx.doi.org/10.1088/0004-637X/777/1/61}{\JournalTitle{\apj}, 777, 61}

\bibitem[{{in 't Zand} {et~al.}(1998){in 't Zand}, {Baykal}, \& {Strohmayer}}]{intZand_1998}
{in 't Zand}, J.~J.~M., {Baykal}, A., \& {Strohmayer}, T.~E. 1998, \href{http://dx.doi.org/10.1086/305362}{\JournalTitle{\apj}, 496, 386}

\bibitem[{{Kislat} {et~al.}(2015){Kislat}, {Clark}, {Beilicke}, \& {Krawczynski}}]{Kislat_2015}
{Kislat}, F., {Clark}, B., {Beilicke}, M., \& {Krawczynski}, H. 2015, \href{http://dx.doi.org/10.1016/j.astropartphys.2015.02.007}{\JournalTitle{Astroparticle Physics}, 68, 45}

\bibitem[{{Lai} \& {Ho}(2003)}]{Lai_2003}
{Lai}, D., \& {Ho}, W.~C. 2003, \href{http://dx.doi.org/10.1103/PhysRevLett.91.071101}{\JournalTitle{\prl}, 91, 071101}

\bibitem[{{Lamb} {et~al.}(1973){Lamb}, {Pethick}, \& {Pines}}]{Lamb_1973}
{Lamb}, F.~K., {Pethick}, C.~J., \& {Pines}, D. 1973, \href{http://dx.doi.org/10.1086/152325}{\JournalTitle{\apj}, 184, 271}

\bibitem[{{Loktev} {et~al.}(2025){Loktev}, {Forsblom}, {Tsygankov}, {Poutanen}, {Mushtukov}, {Di Marco}, {Heyl}, {Kelly}, {La Monaca}, {Ng}, {Ravi}, {Salganik}, {Santangelo}, {Suleimanov}, \& {Zane}}]{Loktev_2025_4U1538}
{Loktev}, V., {Forsblom}, S.~V., {Tsygankov}, S.~S., {et~al.} 2025, \href{http://dx.doi.org/10.1051/0004-6361/202554151}{\JournalTitle{\aap}, 698, A22}

\bibitem[{{Makishima} {et~al.}(1984){Makishima}, {Kawai}, {Koyama}, {Shibazaki}, {Nagase}, \& {Nakagawa}}]{Makishima_1984}
{Makishima}, K., {Kawai}, N., {Koyama}, K., {et~al.} 1984, \JournalTitle{\pasj}, 36, 679

\bibitem[{{Marshall} \& {Ricketts}(1980)}]{Marshall_1980}
{Marshall}, N., \& {Ricketts}, M.~J. 1980, \href{http://dx.doi.org/10.1093/mnras/193.1.7P}{\JournalTitle{\mnras}, 193, 7P}

\bibitem[{{Matsuoka} {et~al.}(2009){Matsuoka}, {Kawasaki}, {Ueno}, {Tomida}, {Kohama}, {Suzuki}, {Adachi}, {Ishikawa}, {Mihara}, {Sugizaki}, {Isobe}, {Nakagawa}, {Tsunemi}, {Miyata}, {Kawai}, {Kataoka}, {Morii}, {Yoshida}, {Negoro}, {Nakajima}, {Ueda}, {Chujo}, {Yamaoka}, {Yamazaki}, {Nakahira}, {You}, {Ishiwata}, {Miyoshi}, {Eguchi}, {Hiroi}, {Katayama}, \& {Ebisawa}}]{Matsuoka_2009_MAXI}
{Matsuoka}, M., {Kawasaki}, K., {Ueno}, S., {et~al.} 2009, \href{http://dx.doi.org/10.1093/pasj/61.5.999}{\JournalTitle{\pasj}, 61, 999}

\bibitem[{{Meszaros} {et~al.}(1988){Meszaros}, {Novick}, {Szentgyorgyi}, {Chanan}, \& {Weisskopf}}]{Meszaros_1988_RVM}
{Meszaros}, P., {Novick}, R., {Szentgyorgyi}, A., {Chanan}, G.~A., \& {Weisskopf}, M.~C. 1988, \href{http://dx.doi.org/10.1086/165962}{\JournalTitle{\apj}, 324, 1056}

\bibitem[{{Mihara}(1995)}]{Mihara_1995}
{Mihara}, T. 1995, PhD thesis, University of Tokyo

\bibitem[{{Mukerjee} {et~al.}(2001){Mukerjee}, {Agrawal}, {Paul}, {Rao}, {Yadav}, {Seetha}, \& {Kasturirangan}}]{Mukerjee_2001}
{Mukerjee}, K., {Agrawal}, P.~C., {Paul}, B., {et~al.} 2001, \href{http://dx.doi.org/10.1086/318655}{\JournalTitle{\apj}, 548, 368}

\bibitem[{{Mushtukov} \& {Tsygankov}(2024)}]{Mushtukov_Tsygankov_2024_Review}
{Mushtukov}, A., \& {Tsygankov}, S. 2024, {Accreting strongly magnetised neutron stars: X-ray Pulsars}, ed. C.~{Bambi} \& A.~{Santangelo} (Singapore: Springer), 4105

\bibitem[{{Mushtukov} {et~al.}(2023){Mushtukov}, {Tsygankov}, {Poutanen}, {Doroshenko}, {Salganik}, {Costa}, {Marco}, {Heyl}, {Monaca}, {Lutovinov}, {Mereminsky}, {Papitto}, {Semena}, {Shtykovsky}, {Suleimanov}, {Forsblom}, {Gonz{\'a}lez-Caniulef}, {Malacaria}, {Sunyaev}, {Agudo}, {Antonelli}, {Bachetti}, {Baldini}, {Baumgartner}, {Bellazzini}, {Bianchi}, {Bongiorno}, {Bonino}, {Brez}, {Bucciantini}, {Capitanio}, {Castellano}, {Cavazzuti}, {Chen}, {Ciprini}, {De Rosa}, {Del Monte}, {Gesu}, {Lalla}, {Donnarumma}, {Dov{\v{c}}iak}, {Ehlert}, {Enoto}, {Evangelista}, {Fabiani}, {Ferrazzoli}, {Garcia}, {Gunji}, {Hayashida}, {Iwakiri}, {Jorstad}, {Kaaret}, {Karas}, {Kislat}, {Kitaguchi}, {Kolodziejczak}, {Krawczynski}, {Latronico}, {Liodakis}, {Maldera}, {Manfreda}, {Marin}, {Marscher}, {Marshall}, {Massaro}, {Matt}, {Mitsuishi}, {Mizuno}, {Muleri}, {Negro}, {Ng}, {O'Dell}, {Omodei}, {Oppedisano}, {Pavlov}, {Peirson}, {Perri}, {Pesce-Rollins}, {Petrucci}, {Pilia}, {Possenti}, {Puccetti}, {Ramsey}, {Rankin},
  {Ratheesh}, {Roberts}, {Romani}, {Sgr{\`o}}, {Slane}, {Soffitta}, {Spandre}, {Swartz}, {Tamagawa}, {Tavecchio}, {Taverna}, {Tawara}, {Tennant}, {Thomas}, {Tombesi}, {Trois}, {Turolla}, {Vink}, {Weisskopf}, {Wu}, {Xie}, \& {Zane}}]{Mushtukov_2023_XPersei}
{Mushtukov}, A.~A., {Tsygankov}, S.~S., {Poutanen}, J., {et~al.} 2023, \href{http://dx.doi.org/10.1093/mnras/stad1961}{\JournalTitle{\mnras}, 524, 2004}

\bibitem[{{Naghizadeh-Khouei} \& {Clarke}(1993)}]{Naghizadeh-Khouei_1993}
{Naghizadeh-Khouei}, J., \& {Clarke}, D. 1993, \JournalTitle{\aap}, 274, 968

\bibitem[{{Nespoli} {et~al.}(2008){Nespoli}, {Fabregat}, \& {Mennickent}}]{Nespoli_2008}
{Nespoli}, E., {Fabregat}, J., \& {Mennickent}, R.~E. 2008, \href{http://dx.doi.org/10.1051/0004-6361:200809645}{\JournalTitle{\aap}, 486, 911}

\bibitem[{{Poutanen}(2020)}]{Poutanen_2020_RVM}
{Poutanen}, J. 2020, \href{http://dx.doi.org/10.1051/0004-6361/202038689}{\JournalTitle{\aap}, 641, A166}

\bibitem[{{Poutanen} {et~al.}(2024){Poutanen}, {Tsygankov}, \& {Forsblom}}]{Poutanen_2024_Galaxies}
{Poutanen}, J., {Tsygankov}, S.~S., \& {Forsblom}, S.~V. 2024, \href{http://dx.doi.org/10.3390/galaxies12040046}{\JournalTitle{Galaxies}, 12, 46}

\bibitem[{{Radhakrishnan} \& {Cooke}(1969)}]{Radhakrishnan_1969_RVM}
{Radhakrishnan}, V., \& {Cooke}, D.~J. 1969, \JournalTitle{\aplett}, 3, 225

\bibitem[{{Rivers} {et~al.}(2010){Rivers}, {Markowitz}, {Pottschmidt}, {Roth}, {Barrag{\'a}n}, {F{\"u}rst}, {Suchy}, {Kreykenbohm}, {Wilms}, \& {Rothschild}}]{Rivers_2010}
{Rivers}, E., {Markowitz}, A., {Pottschmidt}, K., {et~al.} 2010, \href{http://dx.doi.org/10.1088/0004-637X/709/1/179}{\JournalTitle{\apj}, 709, 179}

\bibitem[{{Soffitta} {et~al.}(2021){Soffitta}, {Baldini}, {Bellazzini}, {Costa}, {Latronico}, {Muleri}, {Del Monte}, {Fabiani}, {Minuti}, {Pinchera}, {Sgro'}, {Spandre}, {Trois}, {Amici}, {Andersson}, {Attina'}, {Bachetti}, {Barbanera}, {Borotto}, {Brez}, {Brienza}, {Caporale}, {Cardelli}, {Carpentiero}, {Castellano}, {Castronuovo}, {Cavalli}, {Cavazzuti}, {Ceccanti}, {Centrone}, {Ciprini}, {Citraro}, {D'Amico}, {D'Alba}, {Di Cosimo}, {Di Lalla}, {Di Marco}, {Di Persio}, {Donnarumma}, {Evangelista}, {Ferrazzoli}, {Hayato}, {Kitaguchi}, {La Monaca}, {Lefevre}, {Loffredo}, {Lorenzi}, {Lucchesi}, {Magazzu}, {Maldera}, {Manfreda}, {Mangraviti}, {Marengo}, {Matt}, {Mereu}, {Morbidini}, {Mosti}, {Nakano}, {Nasimi}, {Negri}, {Nenonen}, {Nuti}, {Orsini}, {Perri}, {Pesce-Rollins}, {Piazzolla}, {Pilia}, {Profeti}, {Puccetti}, {Rankin}, {Ratheesh}, {Rubini}, {Santoli}, {Sarra}, {Scalise}, {Sciortino}, {Tamagawa}, {Tardiola}, {Tobia}, {Vimercati}, \& {Xie}}]{Soffitta_2021}
{Soffitta}, P., {Baldini}, L., {Bellazzini}, R., {et~al.} 2021, \href{http://dx.doi.org/10.3847/1538-3881/ac19b0}{\JournalTitle{\aj}, 162, 208}

\bibitem[{{Tobrej} {et~al.}(2023){Tobrej}, {Rai}, {Ghising}, {Tamang}, \& {Paul}}]{Tobrej_2023}
{Tobrej}, M., {Rai}, B., {Ghising}, M., {Tamang}, R., \& {Paul}, B.~C. 2023, \href{http://dx.doi.org/10.1093/mnras/stac3203}{\JournalTitle{\mnras}, 518, 4861}

\bibitem[{{Tsygankov} {et~al.}(2022){Tsygankov}, {Doroshenko}, {Poutanen}, {Heyl}, {Mushtukov}, {Caiazzo}, {Di Marco}, {Forsblom}, {Gonz{\'a}lez-Caniulef}, {Klawin}, {La Monaca}, {Malacaria}, {Marshall}, {Muleri}, {Ng}, {Suleimanov}, {Sunyaev}, {Turolla}, {Agudo}, {Antonelli}, {Bachetti}, {Baldini}, {Baumgartner}, {Bellazzini}, {Bianchi}, {Bongiorno}, {Bonino}, {Brez}, {Bucciantini}, {Capitanio}, {Castellano}, {Cavazzuti}, {Ciprini}, {Costa}, {De Rosa}, {Del Monte}, {Di Gesu}, {Di Lalla}, {Donnarumma}, {Dov{\v{c}}iak}, {Ehlert}, {Enoto}, {Evangelista}, {Fabiani}, {Ferrazzoli}, {Garcia}, {Gunji}, {Hayashida}, {Iwakiri}, {Jorstad}, {Karas}, {Kitaguchi}, {Kolodziejczak}, {Krawczynski}, {Latronico}, {Liodakis}, {Maldera}, {Manfreda}, {Marin}, {Marinucci}, {Marscher}, {Matt}, {Mitsuishi}, {Mizuno}, {Ng}, {O'Dell}, {Omodei}, {Oppedisano}, {Papitto}, {Pavlov}, {Peirson}, {Perri}, {Pesce-Rollins}, {Petrucci}, {Pilia}, {Possenti}, {Puccetti}, {Ramsey}, {Rankin}, {Ratheesh}, {Romani}, {Sgr{\`o}}, {Slane}, {Soffitta},
  {Spandre}, {Tamagawa}, {Tavecchio}, {Taverna}, {Tawara}, {Tennant}, {Thomas}, {Tombesi}, {Trois}, {Vink}, {Weisskopf}, {Wu}, {Xie}, {Zane}, \& {IXPE Collaboration}}]{Tsygankov_2022_CenX3}
{Tsygankov}, S.~S., {Doroshenko}, V., {Poutanen}, J., {et~al.} 2022, \href{http://dx.doi.org/10.3847/2041-8213/aca486}{\JournalTitle{\apjl}, 941, L14}

\bibitem[{{van Kerkwijk} {et~al.}(1989){van Kerkwijk}, {van Oijen}, \& {van den Heuvel}}]{vanKerkwijk_1989}
{van Kerkwijk}, M.~H., {van Oijen}, J.~G.~J., \& {van den Heuvel}, E.~P.~J. 1989, \JournalTitle{\aap}, 209, 173

\bibitem[{{Varun} {et~al.}(2019){Varun}, {Pradhan}, {Maitra}, {Raichur}, \& {Paul}}]{Varun_2019}
{Varun}, {Pradhan}, P., {Maitra}, C., {Raichur}, H., \& {Paul}, B. 2019, \href{http://dx.doi.org/10.3847/1538-4357/ab2763}{\JournalTitle{\apj}, 880, 61}

\bibitem[{{Verner} {et~al.}(1996){Verner}, {Ferland}, {Korista}, \& {Yakovlev}}]{Verner_1996}
{Verner}, D.~A., {Ferland}, G.~J., {Korista}, K.~T., \& {Yakovlev}, D.~G. 1996, \href{http://dx.doi.org/10.1086/177435}{\JournalTitle{\apj}, 465, 487}

\bibitem[{{Weisskopf} {et~al.}(2022){Weisskopf}, {Soffitta}, {Baldini}, {Ramsey}, {O'Dell}, {Romani}, {Matt}, {Deininger}, {Baumgartner}, {Bellazzini}, {Costa}, {Kolodziejczak}, {Latronico}, {Marshall}, {Muleri}, {Bongiorno}, {Tennant}, {Bucciantini}, {Dovciak}, {Marin}, {Marscher}, {Poutanen}, {Slane}, {Turolla}, {Kalinowski}, {Di Marco}, {Fabiani}, {Minuti}, {La Monaca}, {Pinchera}, {Rankin}, {Sgro'}, {Trois}, {Xie}, {Alexander}, {Allen}, {Amici}, {Andersen}, {Antonelli}, {Antoniak}, {Attina'}, {Barbanera}, {Bachetti}, {Baggett}, {Bladt}, {Brez}, {Bonino}, {Boree}, {Borotto}, {Breeding}, {Brienza}, {Bygott}, {Caporale}, {Cardelli}, {Carpentiero}, {Castellano}, {Castronuovo}, {Cavalli}, {Cavazzuti}, {Ceccanti}, {Centrone}, {Citraro}, {D' Amico}, {D'Alba}, {Di Gesu}, {Del Monte}, {Dietz}, {Di Lalla}, {Di Persio}, {Dolan}, {Donnarumma}, {Evangelista}, {Ferrant}, {Ferrazzoli}, {Ferrie}, {Footdale}, {Forsyth}, {Foster}, {Garelick}, {Gunji}, {Gurnee}, {Head}, {Hibbard}, {Johnson}, {Kelly}, {Kilaru}, {Lefevre},
  {Le Roy}, {Loffredo}, {Lorenzi}, {Lucchesi}, {Maddox}, {Magazzu}, {Maldera}, {Manfreda}, {Mangraviti}, {Marengo}, {Marrocchesi}, {Massaro}, {Mauger}, {McCracken}, {McEachen}, {Mize}, {Mereu}, {Mitchell}, {Mitsuishi}, {Morbidini}, {Mosti}, {Nasimi}, {Negri}, {Negro}, {Nguyen}, {Nitschke}, {Nuti}, {Onizuka}, {Oppedisano}, {Orsini}, {Osborne}, {Pacheco}, {Paggi}, {Painter}, {Pavelitz}, {Pentz}, {Piazzolla}, {Perri}, {Pesce-Rollins}, {Peterson}, {Pilia}, {Profeti}, {Puccetti}, {Ranganathan}, {Ratheesh}, {Reedy}, {Root}, {Rubini}, {Ruswick}, {Sanchez}, {Sarra}, {Santoli}, {Scalise}, {Sciortino}, {Schroeder}, {Seek}, {Sosdian}, {Spandre}, {Speegle}, {Tamagawa}, {Tardiola}, {Tobia}, {Thomas}, {Valerie}, {Vimercati}, {Walden}, {Weddendorf}, {Wedmore}, {Welch}, {Zanetti}, \& {Zanetti}}]{Weisskopf_2022}
{Weisskopf}, M.~C., {Soffitta}, P., {Baldini}, L., {et~al.} 2022, \href{http://dx.doi.org/10.1117/1.JATIS.8.2.026002}{\JournalTitle{JATIS}, 8, 026002}

\bibitem[{{Wilms} {et~al.}(2000){Wilms}, {Allen}, \& {McCray}}]{Wilms_2000}
{Wilms}, J., {Allen}, A., \& {McCray}, R. 2000, \href{http://dx.doi.org/10.1086/317016}{\JournalTitle{\apj}, 542, 914}

\end{thebibliography}
\end{document}